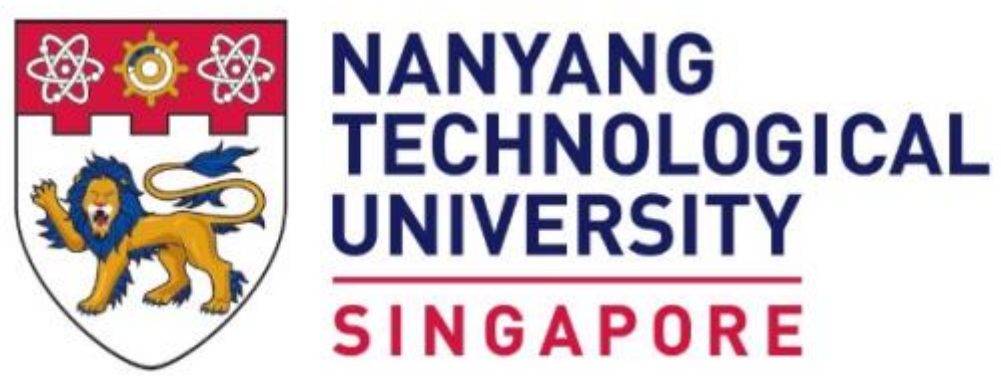

Nanyang Business School

# Reducing Prescription Errors Through Information Intervention: A Field Experiment in Healthcare Operations

Xiaodan Shao
Nanyang Business School, Nanyang Technological University

Vivek Choudhary
Nanyang Business School, Nanyang Technological University

Arnab Majumdar
HealthPlix

**Reducing Prescription Errors Through Information Intervention:**

**A Field Experiment in Healthcare Operations**

Xiaodan Shao (xiaodan001@e.ntu.edu.sg)
Vivek Choudhary (vivek.choudhary@ntu.edu.sg)
Arnab Majumdar

*Abstract*

**Problem definition:** Drug-to-drug interaction (DDI) errors, where two prescribed drugs negatively interact, pose serious risks to patient safety and remain a persistent challenge for clinical decision-making. Existing systems requiring response that are intended to prevent such errors often disrupt doctors' workflow and are frequently overridden. Prior literature suggests that information interventions support better decisions without forcing an immediate response but may also increase cognitive load and potentially errors. This paper examines whether such interventions can reduce DDI errors and foster learning.

**Methodology:** We use a difference-in-differences model to analyze data from a randomized field experiment with the largest electronic medical record platform in India, encompassing 2.81 million prescriptions from ~1,700 doctors. We compare the DDI errors between doctors in the treatment group, who are exposed to the intervention that highlights real-time errors without requiring doctors' responses, and those in the control group without such an intervention.

**Results:** Treatment doctors prescribe 8.6% fewer errors, yielding an estimated annual savings of US$4.8 million in hospitalization costs and potentially saving ~134 lives. Error reduction operates through two mechanisms: *reactive correction*, where doctors remove flagged errors, and *proactive learning*, where they avoid prescription errors upfront. While initial reduction is driven by reactive correction, doctors gradually internalize the information and avoid errors, demonstrating learning. Doctors also become less likely to repeat the same errors and show spillover effects by reducing new errors, suggesting that learning generalizes beyond specific error pairs. The intervention's effectiveness is consistent across doctor types and does not compromise productivity or care quality.

**Managerial implications:** The non-mandatory information intervention can enhance patient safety by reducing prescription errors through proactive learning. By improving decision quality without adding cognitive burden, it provides a scalable and cost-effective approach for healthcare and other operational contexts where risks arise from harmful pairings.

---

## 1. Introduction

Medication errors are a leading cause of avoidable harm in healthcare systems worldwide, with an estimated annual economic burden of US$42 billion (Donaldson et al. 2017). Among these errors, drug-to-drug interactions (DDIs) are particularly critical, where the concurrent use of interacting drugs can cause severe, even fatal reactions (Conti et al. 2022). Although a small fraction of DDIs may be clinically justified, these errors are generally preventable, especially those arising from inadvertent prescribing by doctors (Donaldson et al. 2017). DDI errors are thus a significant operational risk within healthcare systems, as they can lead to increased clinic revisits and hospital admissions, prolonged inpatient stays, and higher treatment costs, ultimately straining healthcare capacity and financial resources (Hughes et al. 2024). Nearly 1% of hospital admissions are attributed to DDIs (Mousavi and Ghanbari 2017). Reducing these errors, therefore, is central to the ongoing effort led by The World Health Organization, in collaboration with global leaders and industry stakeholders (Donaldson et al. 2017).

While DDIs are a global concern in prescription safety, their impact is particularly pronounced in developing countries. In India, medication errors are estimated to have caused 17.46 million deaths over the past decade, with DDIs ranked as the fourth leading cause of mortality (Mude and Mohite 2023). The prevalence of DDI errors in medical departments can reach up to 20%, considerably higher than rates in many developed countries (Ahmad 2015). This scale highlights both the urgency and the opportunity for solutions that can improve prescription safety and mitigate operational risks.

To address DDI errors, doctors are equipped with mandatory alert systems during prescribing. They typically present pop-up alerts that block doctors' interface, requiring justifications before proceeding (Cho et al. 2019). However, these systems have not delivered their expected benefits. Extensive research indicates a high override rate up to 95%, with doctors frequently entering blank or arbitrary justifications, indicating their discontent with workflow interruptions and alert fatigue (Van De Sijpe et al. 2022). Consequently, there is growing interest in alternative approaches that maintain clinical autonomy without compromising productivity, while supporting safe prescribing, particularly under conditions of high patient demand (Joshi et al. 2016).

One such promising avenue for addressing these errors is non-mandatory information interventions, which aim to influence decision-making by providing timely, relevant information without imposing rigid constraints (Karmarkar and Apte 2007). Prior studies in healthcare and operations management suggest that such interventions can promote desirable behaviors (Cadario and Chandon 2020, Choudhary et al. 2022). For instance, reminder-based interventions have been shown to improve medication adherence (Horne et al. 2022), and text-based interventions have been demonstrated to increase influenza vaccination rates (Milkman et al. 2021). However, the evidence remains mixed in their impact. In healthcare, even innocuous information can overload cognition and impair decision quality. For instance, information that induces excessive cognitive demands during hospital admissions can lead doctors to commit more errors and reject patients inappropriately (Kim et al. 2020). Thus, while non-mandatory information interventions may avoid workflow disruptions to reduce errors, they may also introduce unintended cognitive burden potentially triggering new errors. This tension highlights the need for context-specific investigations addressing two critical research questions: First, do such interventions effectively reduce DDI errors? Second, if so, is the reduction driven by proactive behavior (learning) or merely by reactive behavior (corrections of flagged errors)?

To address this gap, we leverage a large-scale randomized field experiment in collaboration with firm HealthPlix, the largest electronic medical records (EMR) system for doctors in India. Approximately 1,700 doctors, spanning a wide range of specialties, were randomly assigned to either a treatment or a control group. Doctors in the treatment group received real-time, non-mandatory interventions highlighting potential DDIs, while those in the control group continued using the system without any change. Using a difference-in-differences (DID) approach on nearly 2.81 million prescriptions, we examine the effects of the intervention on prescription errors and the learning behavior of doctors.

Our findings show that the information intervention reduces errors by nearly 8.6% among treatment doctors compared to the control doctors. Importantly, the results suggest that the reduction is not only driven by reactive correction, but that learning also contributes. Doctors proactively avoid errors in subsequent prescriptions as they internalize the information over time. We find reductions in total, repeated, and new errors even in prescriptions without any reactive corrections, alongside an increase in error-free prescriptions, indicating that the intervention promotes proactive learning rather than producing only immediate corrections. This learning effect builds gradually, emerging a few days after the intervention was introduced. Doctors update their prescribing rules about which pairs trigger errors and avoid previously highlighted harmful combinations, and internalize safer use of interacting molecules. Additional analyses reveal no significant heterogeneity in the effect across doctors' gender, specialty, or workload. Moreover, the intervention does not impede doctors' productivity measured in terms of patient volume and prescription duration, nor does it affect the quality of care as indicated by patients' revisit rates.

This study contributes to the healthcare operations and information intervention literature in three ways. First, it provides rigorous evidence on the effectiveness of non-mandatory information interventions in real-world clinical practice. Prior research has documented mixed and context-dependent outcomes, from substantial behavioral improvements to negligible or even adverse effects in response to information interventions (Hydari et al. 2019, Kim et al. 2020, Milkman et al. 2021, Horne et al. 2022). By leveraging field data from practicing doctors, we show robust evidence of how such interventions shape clinical decision-making.

Second, we show that the information intervention shapes learning by altering how doctors acquire and apply knowledge in their prescribing decisions. Exposure to error information enables doctors to internalize feedback and sustain improvements over time. Unlike surgical tasks, which is rooted in motor skills and repetition (Pisano et al. 2001), prescribing is primarily cognitive, requiring analytical reasoning and diagnostic accuracy (KC and Staats 2012). Our study contributes new evidence documenting learning in this cognitively driven context, broadening healthcare research beyond procedural domains.

Third, we uncover positive spillover effects, as exposure to the intervention not only reduces repeated errors but also lowers the incidence of new errors. This pattern suggests that doctors internalize the underlying principles conveyed by the intervention and apply them proactively across a broader range of prescribing decisions, extending benefits beyond the initially targeted issues.

In summary, our study responds to a critical healthcare challenge by examining whether and how information interventions can reduce prescription errors in a real-world clinical setting. By combining experimental rigor with field conditions, we generate new insights for the literatures on healthcare operations, learning, and information intervention. The findings have direct implications for healthcare policymakers and practitioners seeking practical ways to improve prescription safety. More broadly, our findings on how interventions shape decision-making and learning extend beyond healthcare, providing insights in domains where reducing incompatible pairing errors is critical and human judgment is central.

## 2. Literature Review

Our study builds on research on information interventions on behavior across economics, psychology, and operations management. In particular, we focus on studies examining how such interventions influence decision-making and learning. Given the healthcare context of our study, we also draw on the literature on interaction errors, which explores their prevalence, underlying causes, and the effectiveness of mitigation strategies. Bringing these streams together, we develop a theoretical framework that synthesizes how information intervention can trigger immediate reactive corrections and foster sustained learning, leveraging the framework to guide our empirical analysis.

### 2.1. Information Interventions

Our intervention is grounded in the concept of information interventions, which influence people's behavior through making information salient and highlighting the appropriate action (Song et al. 2018, Rhee et al. 2023, Gao and Tavoni 2024). In clinical decision-support tools, these interventions are broadly categorized into mandatory or non-mandatory (McGreevey et al. 2020). Mandatory alerts are typically interruptive pop-ups or hard stops that force doctors to correct the prescription or justify overrides before proceeding, which have been found not to be very effective in practice (McGreevey et al. 2020). Prior literature shows that mandatory alerts are designed to enforce compliance: they require a correction or justification before the physician can proceed. These hard stops interrupt the prescribing process, fragment attention, add cognitive load, and contribute to alert fatigue, which often lead to superficial compliance or habitual overriding (Phansalkar et al. 2013, Ancker et al. 2017).

Non-mandatory interventions, in contrast, are passive inline prompts that can be ignored, allowing workflow to continue without disruption. Our work is directly related to the stream of non-mandatory information interventions and operates in real time, differing from mandatory alerts in two key aspects. First, it is non-interruptive: it provides real-time information that raises awareness but allows doctors to continue prescribing without disruption. Second, it enables selective engagement: doctors can decide whether and when to act on the intervention, concentrating on the most important cases rather than being forced coercively into action each time. This preserves autonomy and encourages more thoughtful engagement with the suggestions (Phansalkar et al. 2013). Thus, non-mandatory interventions have the potential to be more acceptable and better integrated into doctors' workflows. In our setting, the non-mandatory intervention displays errors progressively within the prescribing interface, replacing forced correction with salient highlights that support meaningful cognitive processing and promote acceptance and learning (Cha et al. 2020).

This design aligns with the operations management literature suggesting that the absence of forced action and presence of information transparency support intrinsic motivation, and promote better user experience, resulting in sustained performance improvement (Argote and Miron-Spektor 2011, Buell et al. 2017). Information-based interventions take various forms, such as feedback, labeling, and reminders (Congiu and Moscati 2022). For instance, Choudhary et al. (2022) find that drivers improve performance when notified about their personal best together with information on their current performance. Similarly, interventions that transparently inform users about delays in a process enhance user experience and reduce complaints (Buell et al. 2017, Yu et al. 2021).

In healthcare operations, however, such interventions have demonstrated mixed outcomes (see Last et al. 2021 for a review). On one hand, they have proven effective in promoting desirable behaviors. For instance, providing information on performance against top-performing peers has improved physician productivity (Song et al. 2018). Similarly, feedback from rejected patients has reduced biases in doctors' admission decisions (Kim and Tong 2023). On the other hand, seemingly innocuous information can impair decision quality due to increased cognitive load, leading doctors to underutilize resources or reduce system efficiency (Sunstein 2017, Kim et al. 2020). Prior research highlights that excessive clinical information, suboptimal data display, and frequent alerting can elevate error rates and undermine patient safety (Nijor et al. 2022). Frequent information exposure can also diminish performance by causing decision-makers to overweight the most recently presented information (Lurie and Swaminathan 2009). Thus, while information interventions are commonly used to promote certain behaviors in operational tasks, the additional information may induce higher cognitive burden and unintended consequences.

In sum, information interventions can yield both benefits and unintended drawbacks, with effects being highly context-dependent. This emphasizes the importance of evaluating not only whether such interventions are effective, but also the magnitude, heterogeneity, and mechanisms through which they operate under real-world field conditions.

### 2.2. Preventing DDI Errors

DDI errors are categorized by severity as *minor, moderate,* or *major*, with *major* ones being the highest clinical priority (Paterno et al. 2009). Minor interactions may slightly increase adverse effects without necessitating therapy changes; moderate interactions can exacerbate patient conditions and may call for adjustments; and major interactions are potentially life-threatening and require immediate medical interventions (Soherwardi et al. 2012). Consistent with prior literature on patient safety from harms of major DDIs, and guidance from health practitioners, the firm focuses the intervention on the most severe *major* interactions.

Extensive research has examined why major DDIs arise and persist in prescriptions. Even well-intentioned doctors may inadvertently prescribe interacting medications due to outdated information, difficulty in finding suitable substitutes, underestimation of interaction severity, or simple human error (Becker et al. 2005). Note that a single drug can contain multiple molecules, and DDIs occur when molecules interact between drugs, so each additional drug in the prescription expands the set of possible molecule-molecule pairings and creates a combinatorial complexity. For instance, Dytor Plus (contains molecules Torasemide and Spironolactone) combined with Telsar (contains molecule Telmisartan) creates a potentially harmful pairing because Spironolactone and Telmisartan interact. As the set of known interactions is large and continually expanding, comprehensive recall is infeasible, especially in settings with short consultation times and heavy patient volumes (Iyer 2017), such as in India, where our study was conducted. As Richens (1975) emphasizes, "*No physician can remember long lists…. Nevertheless, there is a clear need for improving the effectiveness of communication of information about drug interactions…*".

Though a small subset of DDI errors might be clinically justified due to limited drug availability or specific treatment requirements, they are generally considered preventable to avoid serious threats to patient safety (Cho et al. 2019, Van De Sijpe et al. 2022). Approximately half of adverse drug reactions, where DDIs are a common contributor to preventable patient admissions, may be mitigated through early detection during prescribing and better access to interaction information (Van De Sijpe et al. 2022). While prescribing DDIs is not typically considered a criminal offense, adverse outcomes from harmful drug combinations can result in a loss of patient trust, malpractice repercussions, and potential criminal charges against doctors.

Prior literature suggests that physicians frequently override even the most severe 'major' DDI alerts that indicate potentially life-threatening interactions (Weingart et al. 2003). Even when a justification is required, compliance is frequently perfunctory, with doctors entering blank or arbitrary texts just to move forward. Providing these justifications disrupts doctors' workflow and imposes time and attention costs in already constrained consultations (Payne et al. 2015). In addition, information overload exacerbates alert fatigue, where excessive notifications desensitize physicians and reduce responsiveness, ultimately impairing the effect of these alerts. For instance, every additional 100 alerts increases the override rate by 1% (Ancker et al. 2014), highlighting the need to design and test less coercive ways to provide alerts to doctors in reducing errors.

Within healthcare operations management, preventable errors are an increasingly critical but underexplored research area (Kohn et al. 2000, Hodkinson et al. 2020, Ahsani-Estahbanati et al. 2022, Naseralallah et al. 2023). In light of this interest, recent studies have begun to examine the factors that influence the incidence of such errors and inform more effective risk management strategies. For instance, advanced electronic medical records can improve prescription accuracy and reduce medication errors (Hydari et al. 2019). In contrast, operational bottlenecks, including supply disruptions, can heighten the risk of prescribing errors by inducing suboptimal decisions (Park et al. 2025). Despite the promising potential of non-mandatory interventions in reducing prescription errors, there is a lack of field evidence in this broad area. Our study complements this stream by providing field evidence on whether and how non-mandatory information interventions can reduce errors.

### 2.3. Learning and Theoretical Framework

A large body of work in operations management examines how individuals improve performance through learning, emphasizing both tacit learning-by-doing and more deliberate, analytical learning (Adler and Clark 1991, KC and Staats 2012, Choudhary et al. 2022). In the case of prescription writing in our setting, learning is primarily analytical, where doctors actively analyze and process information and update their future decision rules accordingly (Dahlin et al. 2018). In healthcare operations, exposure to errors or near-errors provides rich information for such analytical updating (KC et al. 2013), aligning with the broader error-learning literature that studies how deviations or errors create opportunities for process improvement.

Most existing work on individual learning focuses on procedural and skill-based tasks, such as suturing, requiring hands-on practice, precision, and motor skills (Hatch and Mowery 1998, Pisano et al. 2001, Bavafa and Jónasson 2021, Lapré and Cravey 2022). By contrast, drug prescribing is a cognitive and knowledge-intensive task, involving evaluating symptoms, treatment options, and drug interactions. Our study contributes to this literature by providing evidence on how doctors engage in analytical learning during prescribing, where learning emerges through improvements in a cognitive task environment.

When a doctor writes a prescription that contains a drug interaction, the intervention presents an inline prompt before the prescription is finalized (please refer to Section 3 for details of the interface). The doctor can then choose to remove the drug comprising the interacting molecules or ignore without making any changes. Such interventions increase the salience of errors, prompting doctors to act on the suggestion. By exogenously drawing attention to the risk, the intervention enhances acceptance of corrective changes. Prior literature shows that salient cues can influence behavior, such as red warning signs that deter users from clicking phishing links. Similarly, Allcott and Rogers (2014) found that consumers tend to undervalue long-term gasoline costs compared to upfront purchase prices when selecting automobiles, as other appealing salient features of the vehicle disproportionately influence their decision. More broadly, even when individuals have access to complete information, making certain cues more prominent can reshape decisions (Gao and Tavoni 2024).

**Figure 1. Theoretical Framework**

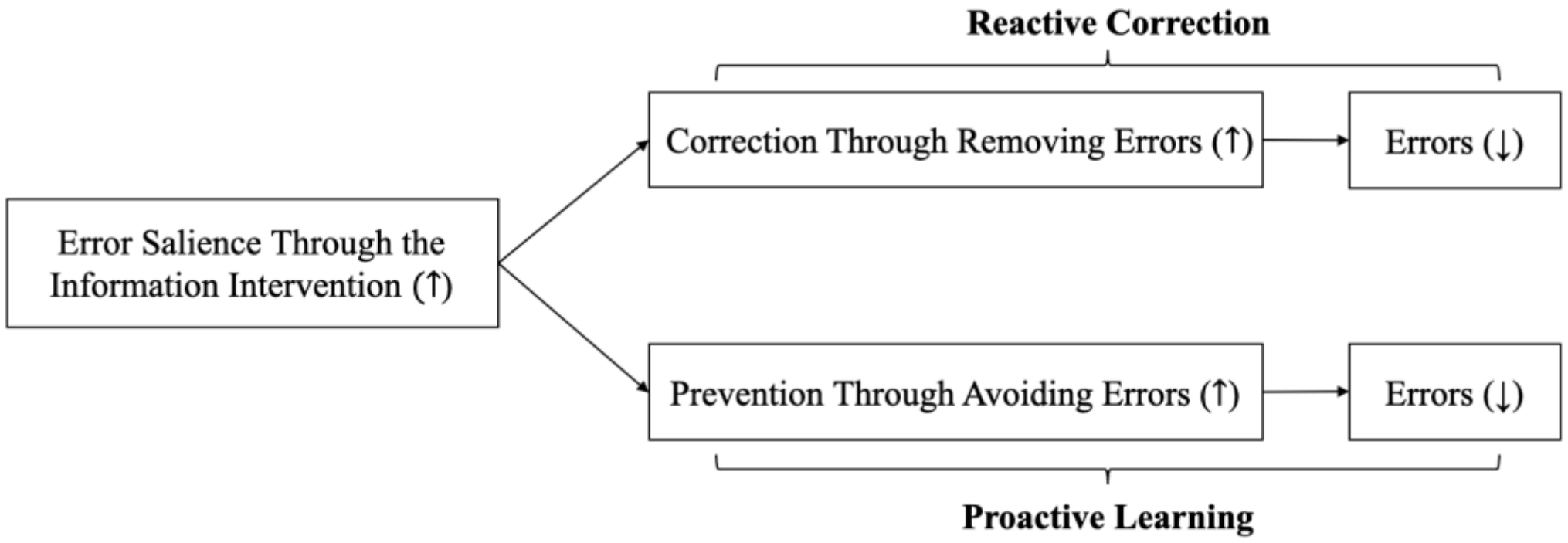

*Note: Reactive correction is indicated by whether the doctor removes the drug(s) causing the interaction. Proactive learning is identified from reductions in errors (e.g., total, repeated, or new) that occur without any correction (metrics used in the empirical analysis).*

When doctors are exposed to the information intervention that highlights errors, their prescribing behavior may change through two potential mechanisms: *reactive correction* and *proactive learning* (Stern et al. 2008, Kim and Xu 2024). Reactive correction refers to the immediate response in which doctors remove drugs that generate the interaction. This correction behavior directly reduces errors through compliance.

By contrast, proactive learning reflects a more enduring shift in behavior through knowledge and belief updating (Kyung et al. 2023). Instead of merely reacting to the intervention, doctors internalize the information provided and proactively avoid prescribing erroneous drug combinations in the first place. Our empirical setting enables us to distinguish between proactive learning and reactive correction by observing whether there is a reduction in errors without correction. We provide the theoretical framework in Figure 1 encapsulating the two mechanisms through which our intervention can reduce errors.

We identify proactive learning using reductions in three key metrics: *total errors, repeated errors, and new errors* in prescriptions without any corrections. While corrections indicate reactive behavior, a reduction in these measures without correction indicates that doctors are systematically improving their decisions, which we analyze in Section 6.

## 3. Experiment Design

In collaboration with the largest health-tech platform in India, the electronic prescription system used by doctors was leveraged to implement a real-time information intervention and examine the doctors' responses and learning behavior.

**Figure 2. Interfaces for Control and Treatment Groups of Doctors**

**Control** **Treatment**

### 3.1. Interface

The platform interface was identical for both the treatment and control doctors before implementing the intervention. The interface used by control doctors remains unchanged throughout the study. The left panel of Figure 2 shows a screenshot of the cloud-based platform that the doctors use to write prescriptions during patient visits before the experiment. On the left panel of the interface (highlighted as 1), doctors can navigate different items listed for each patient, including test results and health trends. At the top (highlighted as 2), the doctor enters the diagnosis for the patient, while the center of the interface (highlighted as 3) is for entering medication details, such as drug names, dosage, frequency, duration, and specific instructions. When a doctor enters the drug name, the molecule composition of the medicine is automatically displayed in parentheses below the drug name (highlighted as 4). After the consultation, the prescription can be handed over to the patient as a digital or a printed copy. The system stores only the final submitted prescriptions and does not capture intermediate versions.

Our intervention for the treatment doctors is an inline prompt in real time that does not disrupt the prescription workflow, as shown in the right panel of Figure 2. The entry of the initial drug molecule (e.g., Methotrexate, highlighted as 5) will turn light red if a subsequently prescribed drug molecule interacts with it (e.g., Rabeprazole, highlighted as 6). In addition, there will be a warning sign next to the later-prescribed interacting molecule (highlighted as 7), where doctors can hover over to check specific interactions. Molecule-level details of these major interactions are displayed at the bottom of the interface (highlighted as 8).

### 3.2. Launch Process and Data Collection

Figure 3 provides the timeline of the experiment. The firm collected baseline data for each doctor from April 4, 2022, to June 29, 2022 (nearly three months). On June 15, the intervention feature was piloted with a small group of doctors to assess usability and identify any technical issues. Feedback from these pilot doctors led to the inclusion of a disclaimer in the interface stating that the error information or warning signs would not appear on the final printed prescription. This adjustment aimed to prevent bias in the results due to doctors' concerns about errors being visible to the patients, ensuring that observed effects were not influenced by potential fear-driven responses.

**Figure 3. Timeline of Intervention Implementation**

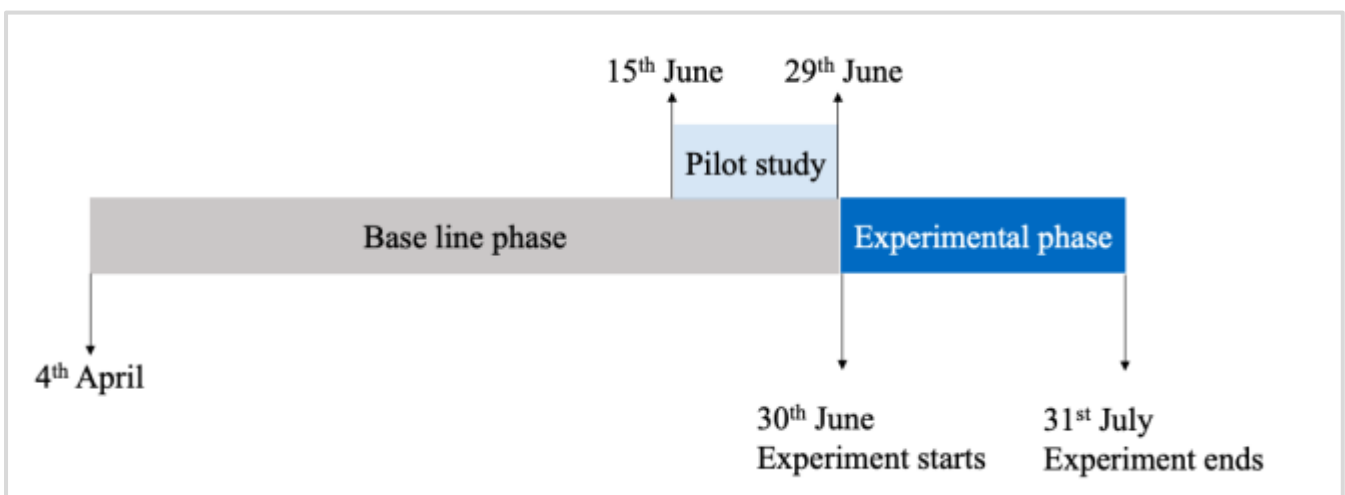

We excluded the pilot doctors to avoid any bias. The intervention was randomly assigned, ensuring that treatment doctors were unaware of the intervention in advance and did not know they were part of an experiment. There was no interaction between treatment and control groups. Further, the changes for treatment doctors were launched simultaneously at the backend of the cloud platform for all the treatment doctors on June 30, 2022. Although the platform highlighted drug interactions, the doctors retained full independence to proceed without responding to the warnings. After running the intervention feature for a month, on July 31, 2022, the firm launched it for all doctors, concluding the experiment[1].

Our collaborator created a proprietary database of major drug interactions from established clinical references that report drug interaction information, including widely-used reliable resources such as Drugs.com listed by U.S. Food and Drug Administration (U.S. FDA 2017). When a doctor prescribes two interacting drugs, the platform cross-references its database and highlights the errors in real-time. For doctors in the treatment group, the interface displays the message, "[Molecule 1] has a major interaction with [Molecule 2]," at the bottom of the screen. This makes clear to doctors that the intervention targets major drug interactions, a concept taught during their medical studies. Pharmacology, a core part of medical school programs such as the Bachelor of Medicine and Bachelor of Surgery, prepares doctors to identify and manage major interactions when prescribing or administering medications. In the pilot phase of the experiment, our collaborator confirmed that doctors recognized the focus of the intervention on major interactions, a point further validated through follow-up interviews.

We have all prescriptions written by doctors, as they typically use a single platform and rarely switch. They evaluate a platform thoroughly before adoption due to the substantial upfront investment in customizing clinical templates, integrating workflows, undergoing training, and managing patient records. Beyond prescription writing, EMR platforms also provide integrated features such as test results management and healthcare analytics, further increasing switching costs. Consequently, once a large platform such as our collaborator's is adopted, doctors tend to remain long-term users. This stability in platform usage ensures that our data captures the full scope of doctors' prescribing behavior, allowing us to observe their responses and learning comprehensively.

### 3.3. Randomization and Variable Balance

The experiment focused on doctors practicing in cities with a significant representation of both general physicians and specialists. Eligible doctors were those who had been active in using the platform for writing prescriptions during the pre-treatment period of baseline collection (April to June 2022), with an average of at least 100 prescriptions per month. This threshold ensured that doctors were regular users and familiar with the prescription system, excluding newcomers and infrequent users. We also excluded specialties where DDIs are unlikely to arise such as ayurvedic doctors, dentists, homeopathy practitioners, and counselors.

To identify the effect of our intervention cleanly, we focus on single-doctor clinics. Multi-doctor environments introduce additional channels, such as peer learning and informal discussions, that could plausibly amplify behavior changes, which requires a separate study as individuals make decisions quite differently from groups (He et al. 2024). As shown in the existing literature, solo-practicing doctors account for a significant portion of the healthcare setup, especially in developing economies with resource-constrained environments (Lilford et al. 2025). In India, due to severe shortages of staff and essential supplies at government health facilities, a substantial share of patients seek care from private providers, which deliver approximately 70% of healthcare services nationwide. Within the outpatient sector, solo practitioners and small independent clinics account for roughly 95% of all providers (Dutta 2019, UKHO 2023). Consistent with these national statistics, approximately 91.4% of clinics on our collaborating platform are single-doctor practices. Therefore, while our design includes solo-doctor clinics for a cleaner identification, the setup remains relevant and highly consistent with the structure of the healthcare system.

Finally, 1,701 doctors were selected from the eligible pool. Roughly one-fourth of them were randomly assigned to the treatment group (473) and the remaining to the control group (1,228), with balance ensured across key characteristics. Doctor attrition was minimal and unrelated to the intervention, as only three doctors from the treatment group who dropped out had stopped prescribing before the experiment began[2].

Our non-identifiable data consists of 2.87 million anonymized prescriptions written by doctors from April 4 to July 31, 2022, covering both baseline and intervention phases. All data entries for doctors and patients were fully anonymized, with no personally identifiable information, ensuring that re-identification was impossible. For doctors, only non-identifiable IDs, specialty, gender, and city were collected, whereas for patients, only anonymized patient IDs linked to each doctor were recorded.

We also remove outliers related to the number of medicines prescribed and the daily number of patients. Specifically, we exclude prescriptions with more than 12 medicines and observations from doctors with an exceptionally high patient volume (more than 113 patients per day). These thresholds were based on the 99th percentile values of the respective variables. Robustness checks in Section A5.1 of Online Appendix, including these outliers, yield results consistent with our main findings (all sections, tables, and figures in the Online Appendix are prefixed with the letter 'A'). Our final dataset consists of 2.81 million prescriptions.

We conduct balance tests to assess the effectiveness of randomization using several variables measured during the pre-treatment period: *daily number of medicines prescribed, total number of active days, total number of patients, daily number of patients* (Bavafa and Jónasson 2021). We also include doctor characteristics such as *number of patient revisits* (proxied for historical care quality; Jain et al. 2024), *experience*, and *specialization* (a binary variable distinguishing general physicians from specialists). For completeness, we additionally compare the dependent variable, *daily number of errors*, between the treatment and control group during the pre-treatment period. Although no doctors were exposed to the DDI intervention before the experiment, the firm can leverage its DDI database to calculate the number of errors per prescription, enabling balance checks and baseline comparisons across groups.

We provide the results of these pre-treatment balance checks in Table 1. Columns (1) and (2) show the means of the variables for the treatment and control groups used for the balance variables and the outcome variable, respectively. Column (3) reports the difference in means between these groups, while columns (4) and (5) provide the corresponding *t*-statistics and *p*-values for these differences. Standard errors are clustered at the doctor level (Donner et al. 1981). All *p*-values exceed 0.05, indicating no statistically significant differences. Additionally, we report standardized differences (*Cohen's d*) in column (6), calculated using the differences in means and the pooled standard deviation. According to Mertens et al. (2022), a *Cohen's d* value below 0.2 is considered a negligible difference. In our analysis, all values are small, with magnitudes around 0.07 and less. To further validate the balance check, we also examine the density and cumulative distribution of these variables in Section A1 of Online Appendix. Visual inspection confirms no significant differences in the distributions between treatment and control groups during the pre-treatment period.

**Table 1. Balance Table for Pre-Treatment Data**

| | (1) | (2) | (3) | (4) | (5) | (6) |
|---|---|---|---|---|---|---|
| **Variables** | **Control** | **Treatment** | **Δ** ***(Treat-Control)*** | ***t*-stat** | ***p*-value** | ***Cohen's d*** |
| *Number of errors* | 2.451<br>(0.125) | 2.156<br>(0.183) | -0.295 | -1.609 | 0.108 | 0.062 |
| *Number of medicines prescribed* | 64.542<br>(1.617) | 61.714<br>(2.980) | -2.828 | -0.949 | 0.343 | 0.044 |
| *Total number of active days* | 49.878<br>(0.186) | 50.365<br>(0.347) | 0.487 | 1.406 | 0.160 | 0.072 |
| *Total number of patients* | 851.446<br>(19.710) | 840.444<br>(37.848) | -11.002 | -0.291 | 0.771 | 0.016 |
| *Number of daily patients* | 16.860<br>(0.363) | 16.637<br>(0.686) | -0.223 | -0.325 | 0.745 | 0.015 |
| *Number of patient revisits* | 1.409<br>(2.632) | 1.456<br>(2.599) | 0.047 | 0.397 | 0.692 | 0.018 |
| *Experience* | 15.879<br>(10.504) | 16.304<br>(10.865) | -0.425 | 0.648 | 0.517 | 0.040 |
| *Specialization (Binary)* | 0.663<br>(0.473) | 0.634<br>(0.482) | 0.028 | -1.066 | 0.287 | 0.059 |

*Note: N = 85,873 for control and N = 33,197 for treatment doctors. Standard errors in parentheses are clustered at the doctor level. Standardized difference (Cohen's d) is calculated as |Δ/pooled standard deviation|. Experience data is available for 80.2% of doctors (N = 69,368 for control and N = 26,433 for treatment doctors), with balanced coverage between control (80.5%) and treatment (79.4%) groups. ***p<0.01, **p<0.05, *p<0.1.*

**Table 2. Summary Statistics**

| Variable | N | Mean | SD | Min | Max |
|---|---|---|---|---|---|
| *Number of errors* | 162,395 | 2.36 | 4.81 | 0 | 121 |
| *Number of medicines* | 162,395 | 65.21 | 65.24 | 0 | 709 |

Table 2 provides summary statistics for the final dataset, which comprises 162,395 daily observations from 1,698 doctors (470 in the treatment group and 1,228 in the control group), encompassing 2.81 million prescriptions. The DDI error rates are comparable to those reported in existing studies (Antoon et al. 2020), with the number of errors per day ranging from 0 to 121 and a mean of 2.36. On average, these doctors prescribe 65 medicines daily.

## 4. Empirical Model

We use the difference-in-differences (DID) model to investigate the effect of the intervention on the number of errors, as specified in equation (1).

$$Errors_{it} = \beta_1 Treatment_i \times After_t + Medicines_{it} + Doctor_i + Date_t + \epsilon_{it} \quad (1)$$

The dependent variable $Errors_{it}$ is the total number of DDI errors prescribed by doctor $i$ on day $t$, where $i$ denotes the individual doctor and $t$ denotes the date. The estimated value of $\beta_1$ represents the effect of the intervention on the number of errors prescribed by the treatment doctors compared to the control doctors. $Treatment_i$ equals 1 for the treatment doctors and zero otherwise. $After_t$ equals 1 for the dates on or after the launch date of the experiment (June 30, 2022) and zero otherwise.

To account for the variation in the number of medicines prescribed, we use $Medicines_{it}$, which is the number of medicines prescribed by doctor $i$ on day $t$. Although controlling for covariates is not essential in a randomized experiment with balanced groups, we include $Medicines_{it}$ as a control to provide precise estimates, given its strong association with errors and in line with standard practice in the medication error literature (Velo and Minuz 2009, Thirumagal et al. 2017, Rasool et al. 2020). We also provide our results without controlling for the *number of medicines* in Section A5.2 of Online Appendix, which are consistent with our main findings. To control for the time-invariant characteristics of doctors, such as age, clinic locations, and specialty, we use the individual-fixed effect $Doctor_i$. Finally, we employ the date-fixed effect $Date_t$ to control for time-varying factors, such as policy changes or seasonal variations.

A parallel trend in the dependent variable between treatment and control doctors prior to the experiment is a key assumption for the validity of the DID estimates. Following the methodology in Cui et al. (2022), we demonstrate the validity of the parallel trend in Section A2 of Online Appendix. We use *Week* dummies to verify that the treatment and control groups statistically exhibit a parallel trend during the pre-treatment period, thereby supporting the valid use of the DID method for estimating the effect of our intervention.

## 5. Results

### 5.1. Model-free Evidence

To provide initial evidence of the intervention effect, we first present model-free comparisons in Table 3. We find that treatment doctors reduce their number of errors by ~8.3% ($= \frac{1.98-2.16}{2.16}$%), while we do not observe any significant changes in errors for the control group. The non-parametric Mann-Whitney U-test supports that the error reduction is statistically significant ($p < 0.01$). This suggests that our intervention has significantly reduced the number of errors prescribed by doctors in the treatment group. We next turn to empirical models to examine the magnitude and robustness of this effect.

**Table 3. Number of Errors Across Groups**

| *Number of errors* | Control Doctor | Treatment Doctor |
|---|---|---|
| Before | 2.45 (5.07, 0.13) | 2.16 (3.84, 0.18) |
| After | 2.46 (5.29, 0.14) | 1.98 (3.87, 0.19) |

*Note: N = 162,395. Standard deviation and standard errors (clustered at the doctor level) are in parentheses.*

### 5.2. Main Results

Column (2) of Table 4 presents the estimation results from equation (1), and shows that the intervention improves prescription safety. The coefficient of $Treatment \times After$ is negative and significant ($\beta_1 = -0.186, p < 0.01$), suggesting that the intervention reduces the errors prescribed by the treatment doctors relative to those by the control doctors. On average, treatment doctors prescribe nearly 8.6% (= 0.186/2.16%) fewer errors due to the intervention, with an average baseline error of 2.16. Extrapolating these results, the estimated reduction in errors translates into annual savings for patients of US$4.8 million in hospitalization costs and potentially saving 134 lives (see detailed calculations in Section A3 of Online Appendix).

**Table 4. Effect of the Intervention on Errors**

| | (1) | (2) | (3) |
|---|---|---|---|
| | | *Number of errors* | |
| *Treatment × After* | | −0.186*** | |
| | | (0.052) | |
| *Treatment* × $Week_1$ | | | −0.200*** |
| | | | (0.055) |
| *Treatment* × $Week_2$ | | | −0.152** |
| | | | (0.067) |
| *Treatment* × $Week_3$ | | | −0.193** |
| | | | (0.075) |
| *Treatment* × $Week_4$ | | | −0.201*** |
| | | | (0.073) |
| *Treatment* × $Week_5$ | | | −0.181** |
| | | | (0.081) |
| *Number of medicines* | 0.043*** | 0.043*** | 0.043*** |
| | (0.002) | (0.002) | (0.002) |
| $R^2$ Adj. | 0.786 | 0.786 | 0.786 |

*Note: N = 162,395. SEs in parentheses are clustered at the doctor level. Estimated with doctor and date fixed effects. * p < 0.1, ** p < 0.05, *** p < 0.01.*

As a robustness check, we also estimate a negative binomial model to account for the count nature of the dependent variable, which shows an 8.9% reduction in errors, closely aligned with the 8.6% reduction in our specification (refer to Section A4.1 of Online Appendix). We further obtain consistent findings using two complementary approaches. The first involves nearest-neighbor 1:1 matching based on pre-treatment characteristics to construct comparable treatment and control groups of similar size. The second applies generalized synthetic control methods to account for time-varying unobserved confounders. Detailed analyses and results for both approaches are provided in Sections A4.2 and A4.3 of Online Appendix, respectively.

Moreover, the reduction is not a short-lived response due to the novelty effect. To investigate the consistency of the intervention effect, we conduct a weekly analysis of the treatment effect, finding that the effect of the intervention appears within individual weeks and remains consistent over time, as shown in column (3) of Table 4. These findings suggest that the impact of the intervention remains stable throughout the entire one-month experiment and is not driven by the initial novelty of the intervention that quickly wears off.

### 5.2.1. Error Reduction Not Driven by Fewer Medicines

**Figure 4. Effect of the Intervention on Number of Errors Through Number of Medicines**

$\alpha_1$
Number of Medicines
$\beta_3$
DDI Intervention
$\beta_2$
Number of Errors

In our main results (Table 4), the number of medicines prescribed is shown as a significant indicator of DDI errors, raising a natural question: do doctors reduce errors by prescribing fewer medicines? As illustrated in Figure 4, the intervention may reduce errors that are mediated through an indirect effect of number of medicines, thereby lowering the likelihood of errors.

Following Altman et al. (2021), we conduct a similar mediation analysis. In contrast to equation (1), we now estimate two separate models: (i) a *mediator* model that estimates the effect of the intervention on the number of medicines, as specified in equation (2); and (ii) an *outcome* model that assesses the impact of the intervention and medicines on the number of errors, as specified in equation (3). The coefficient $\beta_2$ in equation (3) represents the direct effect of the intervention on changes in errors, which captures the portion of the intervention effect that is not mediated by the number of medicines (Hayes and Rockwood 2020). $\alpha_1 \times \beta_3$ captures the indirect effect on errors mediated through the number of medicines.

$$Medicines_{it} = \alpha_1 Treatment_i \times After_t + Doctor_i + Date_t + \mu_{it} \quad (2)$$

$$Errors_{it} = \beta_2 Treatment_i \times After_t + \beta_3 Medicines_{it} + Doctor_i + Date_t + v_{it} \quad (3)$$

Following Hayes and Rockwood (2020), we implement a bootstrapping procedure[3] (Altman et al. 2021) and provide the results in Table 5. Column (1) shows that the intervention does not significantly affect the number of medicines prescribed, as the confidence interval for $\alpha_1$ (-0.665, 2.765) includes zero. In contrast, in column (2), the interval for $\beta_2$ (-0.292, -0.082) suggests a statistically significant reduction in errors due to the treatment, aligning with our main analysis. The confidence interval for the indirect effect $\alpha_1 \times \beta_3$ (-0.028, 0.120) includes zero, indicating no evidence of a mediation effect through the number of medicines.

**Table 5. Mediation Effect of Treatment Through the Number of Medicines Prescribed**

| | (1) | (2) |
|---|---|---|
| | *Number of medicines* | *Number of errors* |
| *Treatment× After* | 1.026 | -0.186*** |
| | (0.872) | (0.053) |
| *Number of medicines* | | 0.043*** |
| | | (0.002) |
| $R^2$ Adj. | 0.739 | 0.786 |

*Note: N = 162,395. SEs in parentheses are bootstrapped and clustered at the doctor level. Estimated with doctor and date fixed effects. * p < 0.1, ** p < 0.05, *** p < 0.01.*

Thus, these results suggest that doctors reduce errors not by prescribing fewer or removing more medicines, but by selecting safer alternatives, demonstrating increased caution even though the quantity of medicines remains unchanged. Therefore, in line with existing literature, including the number of medicines as a control variable does not alter our findings but enhance the precision of our estimates mitigating potential confounding, strengthening the robustness of our main results. We thus include the number of medicines as a control variable for all the analyses henceforth.

## 6. Doctors' Learning Behavior: Evidence from Uncorrected Prescriptions

After establishing both the effectiveness and persistence of our intervention, we turn to how it changes the doctors' prescribing: are the error reduction primarily the result of *reactive correction* (removing flagged interactions) or *proactive learning* (avoiding future errors)? As outlined in our theoretical framework, understanding the channel through which the intervention operates is critical, as it reveals whether the behavioral change stems from temporary compliance or deeper cognitive adaptation, shaping both the theoretical contribution and the practical implications for policy design. While we cannot conduct a separate mechanism experiment that fully disentangles the effects and lack comprehensive within-prescription clickstream, we provide evidence of proactive learning by analyzing prescriptions where reactive correction is absent.

After observing our intervention, a treatment doctor can remove an interacting drug by correcting the prescription before sharing it with the patient, which is captured by the system. Thus, corrections are applicable only to treatment doctors. When no corrections are made, the final prescriptions are the same as the doctor's initial choices, and the corrective behavior is absent in these prescriptions. Therefore, we leverage these uncorrected prescriptions in the treatment group and compare them to all prescriptions in the control group (which, by design, involve no correction) to identify learning. This comparison isolates the effect of reactive correction, and error reductions in the uncorrected prescriptions would indicate that doctors internalize the information and proactively avoid errors through learning.

While learning may also occur in corrected prescriptions (e.g., a doctor might remove one error while simultaneously avoiding another interaction in the same prescription), our focus on uncorrected prescriptions provides a conservative, lower bound estimate of the learning effect, as it captures only cases with no reactive correction.

### 6.1. Overall Learning

Building on the comparison in uncorrected prescriptions between control and treatment doctors, column (1) of Table 6 shows a significant reduction in the total errors (-0.222, $p < 0.01$) due to our intervention. Because this decline arises even when reactive correction is excluded, it strongly suggests that proactive learning is an independent mechanism: doctors avoid errors when prescribing upfront and not by fixing them after being highlighted.

### 6.2. Learning from Errors: Repeated Errors and Spillover Effects

To better understand how the intervention affects prescribing behavior, we decompose total errors into two categories: *repeated errors* and *new errors*. This is consistent with the idea that information-based learning is often measured using one's prior similar failure events, and that repeated errors directly indicate whether doctors are learning (Dahlin et al. 2018). In contrast, new errors indicate whether such learning transfers to new situations, aligning with evidence that exposure to errors can strengthen individuals' ability to cope with and generate solutions to new problems (Dahlin et al. 2018).

We define these errors at a doctor level to accurately capture each doctor's own learning trajectory, whether a doctor continues to make the same errors or avoids repeating them after exposure to the intervention. A $Repeated\ error_{it}$ is an error prescribed by $doctor_i$ on $day_t$ that was previously prescribed by this doctor from April 4, 2022 (start of baseline collection), up to $day_{t-1}$. A $New\ error_{it}$ is an error prescribed by $doctor_i$ on $day_t$ that was previously not prescribed by this doctor from April 4, 2022, up to $day_{t-1}$. For example, on July 15, 2022, repeated errors are identified by comparing prescriptions from that day with all errors prescribed between April 4 and July 14, while new errors are those never previously observed for that doctor. These definitions follow established literature and are applied consistently to errors made by all doctors, irrespective of whether these errors have been flagged by the intervention (Crespin et al. 2010, Rodriguez-Gonzalez et al. 2012, Dahlin et al. 2018).

Table 6 shows a significant reduction in repeated errors within uncorrected prescriptions (-0.190, $p < 0.01$, column 2). This suggests that doctors internalize information from prior flagged interactions and are less likely to repeat the same errors, even when no correction behavior occurs. Such reduction provides strong evidence of learning, as doctors internalize earlier error information and adjust their subsequent prescribing decisions.

While most of the effect is driven by the reduction in repeated errors, we also find a smaller but significant reduction in new errors (-0.032, $p < 0.05$, column 3). This indicates that the intervention's impact extends beyond preventing repeated errors: doctors appear to generalize the information to novel prescribing contexts, demonstrating spillover effects. In other words, our intervention improves the awareness of interaction risks that improve overall prescription quality, not just the avoidance of previously flagged errors.

To ensure robustness of our estimation, we replicate these analyses using a 7-day moving window to classify repeated and new errors. Specifically, repeated errors are defined as those occurring within the previous 7 days for a given doctor, while new errors are those not observed during that period. The results remain consistent with a significant reduction in repeated and new errors, and we provide the robust findings in Section A5.3 of Online Appendix.

Together, these findings contribute to broader literature on learning in operational settings by showing that targeted interventions can trigger both intended effect (avoiding repeated errors) and spillover effect (avoiding new errors). Prior work has often emphasized reactive corrections or compliance-based responses to alerts and protocols (Ancker et al. 2017), but our results highlight the proactive role of information-based learning in shaping sustained improvements in decision quality.

**Table 6. Effect of the Intervention on Uncorrected Prescriptions**

| | (1) | (2) | (3) |
|---|---|---|---|
| | *Number of errors* | *Number of repeated errors* | *Number of new errors* |
| *Treatment × After* | -0.222*** | -0.190*** | -0.032** |
| | (0.051) | (0.049) | (0.015) |
| $R^2$ Adj. | 0.786 | 0.780 | 0.230 |

*Note: N = 162,388. SEs are clustered at the doctor level. Estimated with all control variables, doctor fixed effect, and date fixed effect. * p < 0.1, ** p < 0.05, *** p < 0.01.*

### 6.3. Learning Toward Zero Error: Error-free Prescriptions

The preceding analyses on repeated and new errors illustrate how doctors reduce errors, either by avoiding past errors or by generalizing knowledge to new prescribing situations. To further validate that proactive learning contributes to more than just reducing specific errors, we next examine whether doctors internalize the information sufficiently to eliminate them, producing error-free prescriptions, i.e., prescriptions with zero errors. In uncorrected prescriptions, while reduction in errors may indicate safety improvement, error-free prescriptions require doctors to integrate the information more broadly, consistent with more systematic and internalized learning in the absence of external cues.

We again restrict the analysis to prescriptions that are uncorrected and, importantly, error-free, where doctors avoid any errors to begin with. Because these prescriptions contain no errors, there is no salience from the intervention that could trigger reactive behavior. Any improvement in error-free uncorrected prescriptions therefore indicates proactive behavior, with doctors learning to systematically choose safer combinations ex ante rather than respond to the intervention. Table 7 provides evidence consistent with this proactive behavior. We find that the coefficient of the interaction term *Treatment × After* is positive and statistically significant (0.493, $p < 0.01$), indicating that treatment doctors became more likely to issue prescriptions without any errors following the intervention. Because these prescriptions involve no corrective removal, the increase in error-free prescriptions suggests that doctors do not merely cut down on the number of errors, but consistently prioritize safety, reinforcing the practices learned that lead to systematically safer prescribing.

**Table 7. Effect of the Intervention on Error-free Uncorrected Prescriptions**

| | *Number of error-free prescriptions* |
|---|---|
| *Treatment × After* | 0.493*** |
| | (0.157) |
| $R^2$ Adj. | 0.914 |

*Note: N = 162,388. SEs are clustered at the doctor level. Estimated with all control variables, doctor fixed effect, and date fixed effect. * p < 0.1, ** p < 0.05, *** p < 0.01.*

Taken together with the reduction of repeated and new errors, this analysis reinforces our central finding: the intervention fosters learning beyond the specific targeted issues. Doctors do not merely cut down on the number of errors but also increasingly aim to eliminate errors altogether.

### 6.4. Learning Dynamics

Having established that treatment doctors not only reduce repeated and new errors, but also increasingly write fully error-free prescriptions, we next examine how this learning unfolds over time. Understanding the temporal dynamics allows us to distinguish between immediate effects of the intervention and more gradual, sustained adjustments in prescribing behavior.

To do so, we estimate a series of regressions on uncorrected prescriptions, incrementally expanding the analysis window from the intervention start date by adding one day at a time. Columns (1) to (5) of Table 8 show the treatment effects using data on the first day, the first two days, up to the first five days of the intervention, respectively. This allows us to capture the early trajectory of behavior change. We find that the reduction in errors through learning is not immediate, emerging only after the first few days following the experiment launch (results for day 6, 7, and beyond are consistent and thus omitted for brevity). To test the sustained effect of the intervention, we estimate the effect from the fifth day onward, as reported in Column (6). The significant negative coefficient suggests that from day 5 onward, doctors' learning appears to consolidate, driving a sustained reduction in errors.

This temporal pattern is consistent with a gradual learning process: in the first few days, doctors likely experiment with different molecules to avoid errors, and thereafter they write fewer errors proactively from the outset. In line with this, the reduction in errors during the first week (Table 4) is primarily driven by reactive correction, as proactive learning has not yet fully developed, indicating that the initial trial-and-error phase gives way to a more stable pattern of safer prescribing. This evolution supports our interpretation that doctors take time to internalize the information from the intervention and update their prescribing practices beyond mere reactive correction, leading to enduring improvements in prescribing behavior.

**Table 8. Learning Dynamics on Uncorrected Prescriptions**

| | (1) | (2) | (3) | (4) | (5) | (6) |
|---|---|---|---|---|---|---|
| | *Number of errors* | | | | | |
| | *Day 1* | *Day 2* | *Day3* | *Day 4* | *Day 5* | *Day 5 onwards* |
| *Treatment × After* | 0.204* | -0.019 | -0.124* | -0.108 | -0.180*** | -0.236*** |
| | (0.123) | (0.086) | (0.072) | (0.068) | (0.065) | (0.054) |
| N | 120,575 | 122,060 | 123,462 | 123,979 | 125,442 | 157,479 |
| $R^2$ Adj. | 0.786 | 0.786 | 0.786 | 0.786 | 0.786 | 0.786 |

*Note: SEs in parentheses are clustered at the doctor level. Estimated with all control variables, doctor fixed effect, and date fixed effect. * $p < 0.1$, ** $p < 0.05$, *** $p < 0.01$.*

### 6.5. Learning to Recombine: What Doctors Internalize

Given that DDIs ultimately occur at the molecule-molecule combinations, doctors may become more cautious about specific molecules that are frequently involved in interaction pairs, or they learn the pair-level knowledge about which combination of molecules are harmful and adjust how they assemble molecules. If doctors become "molecule-averse," we should see changes in the frequency of the molecule set they use. If instead they learn "pair knowledge," molecule use should remain stable, but risky co-prescribing patterns should decline. Accordingly, we estimate the intervention effect on uncorrected prescriptions[4], focusing on (i) the total and distinct number of molecules prescribed, and (ii) the total and distinct number of molecules prescribed from the major interaction database. We evaluate whether doctors reduce errors by either excluding molecules appearing in the major interaction database (interaction-set) from prescriptions or avoiding specific molecule pairs that previously caused interactions, in essence, determining whether their learning targets individual molecules or problematic pairs.

Table 9 shows that the intervention does not change doctors' overall use of molecules (total, distinct). Columns (1)-(2) indicate no significant effect on the total or unique number of molecules prescribed. This pattern is consistent with the stable number of medicines in Table 5 and suggests that error reductions are not driven by simply prescribing fewer drugs or concentrating on fewer choices of molecules. Columns (3)-(4) similarly show no significant change in the total or unique number of interaction-set molecules (irrespective of whether they are causing an error or not) indicating that treatment doctors are not reducing errors by narrowing their molecule set but they are systematically changing how they combine molecules.

**Table 9. Effect of the Intervention on Molecules on Uncorrected Prescriptions**

| | (1) | (2) | (3) | (4) | (5) |
|---|---|---|---|---|---|
| | *Total molecules* | *Distinct molecules* | *Total molecules (interaction-set)* | *Distinct molecules (interaction-set)* | *Number of errors* |
| *Treatment × After* | 0.289 | 0.172 | -0.006 | -0.037 | -0.222*** |
| | (0.525) | (0.249) | (0.245) | (0.112) | (0.051) |
| $R^2$ Adj. | 0.980 | 0.890 | 0.982 | 0.895 | 0.786 |

*Note: N = 162,388. SEs are clustered at the doctor level. Estimated with all control variables, doctor fixed effect, and date fixed effect. * p < 0.1, ** p < 0.05, *** p < 0.01.*

Therefore, columns (1)-(4) suggest that doctors do not eliminate molecules, rather they recombine them to reduce errors (reproduced from Table 6) as indicated in column (5). This pattern is consistent with learning-driven recombination: rather than earmarking molecules causing interactions, doctors update their prescribing rules about which pairs trigger errors and selectively avoid pairing molecules in ways that have produced errors. It demonstrates that doctors learn how to use interacting molecules in a more error-free way. It also ties to our early findings in Table 6 about the reductions in repeated and new errors, suggesting that doctors are reorganizing their molecule choices to reduce repeated errors without introducing new interaction pairs in prescriptions. Overall, our intervention helps doctors build pair-level interaction knowledge and learn how to recombine their molecule choices. Consequently, doctors keep a similar set of molecules in use but assemble and deliver them to patients in safer combinations.

## 7. Heterogeneous Treatment Effect

After estimating the overall effect of the DDI intervention and providing evidence of learning, we further analyze the heterogeneity of treatment effect across doctor types, including gender, specialty, and workload. Prior research shows that female doctors tend to spend more time on documentation and administrative tasks (Malacon et al. 2024), which may influence cognitive load and their responsiveness to interventions. Behavioral differences across specialties are also known, with evidence of systematic variation in clinical and professional conduct (Cooper et al. 2024). Lastly, doctor workload has been shown to affect quality of care, with higher patient volumes associated with reduced diligence (Powell et al. 2012), suggesting that learning and behavior change may differ by workload constraints. Our heterogeneity analysis along these dimensions helps determine whether the effects of the intervention are broadly consistent or concentrated among specific subgroups of providers.

For workload heterogeneity, we classify each doctor's daily workload as high or low based on their individual median patient volume observed in the dataset. To better account for the baseline differences in errors across subgroups, we employ negative binomial regressions incorporating patient volume as an offset, effectively normalizing errors per patient visit and adjusting for the greater room to commit errors under higher volume. As shown in Table 10, we do not find statistically significant differences in treatment effects across gender, specialty, or workload. The triple interaction terms are all statistically insignificant. These results indicate that the impact of the intervention on error reduction is broadly consistent across doctor subgroups and is not driven by any particular type of provider.

**Table 10. Effect of the Intervention on Errors Across Doctor Types (Negative Binomial Model)**

| | (1) | (2) | (3) |
|---|---|---|---|
| | *Number of errors* | | |
| *Treatment × After* | -0.102*** | -0.071*** | -0.074*** |
| | (0.029) | (0.023) | (0.023) |
| *Treatment × After × Female* | 0.004 | | |
| | (0.040) | | |
| *Treatment × After × GeneralPhysician* | | -0.071 | |
| | | (0.043) | |
| *Treatment × After × HighWorkload* | | | -0.036 |
| | | | (0.023) |
| $R^2$ Adj. | 0.278 | 0.279 | 0.279 |

*Note: N = 162,395. SEs in parentheses are clustered at the doctor level. Estimated with all control variables, doctor fixed effect, and date fixed effect. * p < 0.1, ** p < 0.05, *** p < 0.01.*

## 8. Effect on Productivity and Quality of Care

One concern that may arise from the intervention is whether it adversely affects doctors' productivity or the quality of care they provide. Our study examines both aspects. To assess productivity, we consider patient volume and time taken to issue prescriptions. For quality of care, we examine the effect on follow-up visits as a proxy for patient outcomes.

### 8.1. Productivity

Our analysis indicates that the intervention does not negatively impact doctors' productivity. We use two measures for productivity: daily patient volume[5] and average prescription duration (Haggag et al. 2017). However, due to data limitations, only timestamps marking the time a prescription is finally saved before printing are available. Prescription duration is therefore derived using the difference between consecutive save times, $Save\ time_{i,k+1}$ - $Save\ time_{i,k}$, where $k$ represents the $k$-th prescription written by the doctor $i$ in a day. This measurement provides an upper-bound estimate of prescription duration (Bavafa and Terwiesch 2019). Table 11 reports that daily patient volume increases slightly by 0.482 ($p < 0.1$) for treatment doctors following the intervention (column 1). Columns (2) and (3) display the mean and median prescription durations, respectively, revealing no significant changes. These findings suggest that the DDI intervention does not substantially reduce doctors' productivity.

**Table 11. Effect of the Intervention on Doctor Productivity and Quality of Care**

| | (1) | (2) | (3) | (4) |
|---|---|---|---|---|
| | Productivity | | | Quality of Care |
| | *Patient volume* | *Prescription duration (mean)* | *Prescription duration (median)* | *Number of 7-day revisit patients* |
| *Treatment × After* | 0.482* | -0.020 | -0.018 | 0.079 |
| | (0.268) | (0.102) | (0.111) | (0.057) |
| N | 162,395 | 148,824 | 148,824 | 162,395 |
| $R^2$ Adj. | 0.718 | 0.395 | 0.366 | 0.684 |

*Note: The prescription durations are available for ~92% of the prescriptions. SEs in parentheses are clustered at the doctor level. Estimated with all control variables, doctor fixed effect, and date fixed effect. * p < 0.1, ** p < 0.05, *** p < 0.01.*

### 8.2. Quality of Care

Due to restricted access to direct patient health outcomes, we use revisit rates as a proxy for patient outcomes. Specifically, we calculate the number of 7-day revisit patients to assess the immediate post-treatment phase. Clinically, a revisit within seven days often indicates that the initial treatment for the patient was insufficient, or early complications have arisen, requiring prompt attention (Orosco et al. 2015).

This timeframe is critical because it captures acute issues from the initial care directly linking to the quality and effectiveness of the treatment. The 7-day revisit metric is widely recognized and utilized across healthcare systems for benchmarking and comparison against established standards (Jain et al. 2024). In our context, this is a particularly good proxy because doctors typically do not charge patients for follow-up visits within seven days (Yengkhom 2023).

As shown in Column (5) of Table 11, there is no significant increase in revisits among patients treated by doctors following the implementation of the intervention, suggesting that the DDI intervention does not adversely impact patient health outcomes. We also evaluate patient outcomes using alternative windows, such as 14-day and 15-day revisits, with consistent results, though these results are omitted for brevity.

## 9. Alternative Explanations for the Results: Platform Experience or Recent Patient Load

We explore and test alternative explanations for the error reduction and rule out these possibilities to ensure the effect is attributable to the intervention. Specifically, we control for dynamic attributes such as doctors' platform experience and their recent patient load (Bavafa and Jónasson 2024), because doctors may improve as they gain familiarity with the EMR platform, or their prescribing behavior may also change when their recent patient demand fluctuates. Platform experience is measured by the cumulative number of active days a doctor has used the platform since the start of the baseline collection, while recent patient load is measured as the number of patients treated in the past week. These variables are included separately and then together in our model specification to assess whether either factor could account for the reduction in errors.

**Table 12. Effect of the Intervention on Errors (with Dynamic Attributes as Controls)**

| | (1) | (2) | (3) | (4) |
|---|---|---|---|---|
| | *Number of errors* | | | |
| *Treatment × After* | −0.186*** | -0.182*** | −0.186*** | -0.182*** |
| | (0.052) | (0.052) | (0.052) | (0.052) |
| *Number of medicines* | 0.043*** | 0.043*** | 0.043*** | 0.043*** |
| | (0.002) | (0.002) | (0.002) | (0.002) |
| *Number of cumulative active days* | | -0.007** | | -0.007** |
| | | (0.003) | | (0.003) |
| *Number of patients in the past week* | | | -0.002 | 0.005 |
| | | | (0.009) | (0.010) |
| $R^2$ Adj. | 0.786 | 0.786 | 0.786 | 0.786 |

*Note: N = 162,395. SEs in parentheses are clustered at the doctor level. Estimated with doctor and date fixed effects. * p < 0.1, ** p < 0.05, *** p < 0.01.*

Columns (2) to (4) of Table 12 shows that the coefficient of *Treatment × After* (-0.182, $p$<0.01) remains statistically significant and similar across specifications. In column (4), where we jointly control for platform experience and recent patient load, the coefficient of *Treatment* × *After* is essentially unchanged relative to earlier models, indicating that the intervention delivers an incremental, information-driven learning effect beyond what doctors would achieve through platform familiarity alone. These results rule out platform experience and short-run demand as alternative explanations for the error reduction.

## 10. Discussion and Conclusion

Our study provides robust evidence that a simple, non-mandatory, and cost-effective information intervention can significantly reduce prescription errors and promote learning. Treatment doctors reduce errors by 8.6% relative to the control, driven by proactive learning beyond reactive correction. Doctors gradually internalize the information, resulting in sustained improvements in prescribing behavior. The effects are consistent across gender, specialty, or workload, suggesting broad applicability across provider types. More importantly, these improvements do not come at the expense of productivity or quality of care.

To isolate the learning from reactive correction, we examine the prescriptions without any corrections. Analysis of uncorrected prescriptions shows that treatment doctors systematically avoid errors, including the decline in repeated or new errors and the increase in error-free prescriptions, consistent with deeper internalization of safe prescribing. This spillover effect aligns with literature that feedback on focal tasks can transfer to new domains and improve overall performance (Argote and Miron-Spektor 2011). More broadly, our findings suggest that such interventions can do more than prevent immediate errors by inducing persistent behavioral change with benefits that extend beyond focal tasks.

Molecule-level analysis further shows what doctors learn and how they improve. Treatment doctors do not reduce errors by narrowing the molecule sets they use or avoiding interaction-potential molecules. Instead, they recombine a similar set of molecules into safer pairings, supporting our theoretical framework in which information salience facilitates internalization and proactive learning. Operationally, the stable molecule set implies that implementing the intervention at a larger scale, doctors and pharmacies may not need to significantly adjust drug inventories or ordering behavior.

These learning effects also help interpret that overrides may reflect avoidable prescribing errors. In our setting, an error is defined as the presence of a DDI in a prescription. For treatment doctors post intervention, these errors represent DDIs that were ignored, since the intervention does not impose any additional requirement during prescribing. Prior research shows that overridden alerts are typically overridden again in future encounters (~99% of cases such as in Ancker et al. 2017), suggesting that overrides may become habitual rather than clinically deliberate. Consistent with this concern, the post-intervention reduction in repeated errors (-0.190 in Table 6, ~11%) indicates that at least some prior overrides were not clinically justified and that intervention exposure supports reassessment and adaptation over time.

Recognizing the effectiveness and operational value of the intervention, the company incorporated it as a permanent feature and rolled it out platform-wide, converting the experimental evidence into practical improvement in prescription safety at scale. In follow-up exchanges, doctors also described the feature as among the most valuable on the platform. After the August 1, 2022, rollout, a comparable control group was no longer available, but we collect additional six weeks of data to provide model-free evidence within the treatment doctors. Section A6 of Online Appendix shows that error reduction remains stable within the treatment group eleven weeks after implementation, consistent with a sticky effect over time.

The finding that learning emerges within a few days also suggests a possible conjecture for settings undergoing digitalization, particularly in developing countries. When real-time interventions are not consistently available due to infrastructure limitations or cost considerations, early exposure to high-quality interventions may still trigger learning that persists. Managers might therefore focus on early-phase infrastructure investment that helps users internalize core principles, after which proactive learning sustains error reduction.

From a managerial perspective, the study highlights the potential of light-touch information interventions to improve decision quality in complex operational environments. The learning mechanism is generalizable beyond prescription errors, because it draws on general decision processes, including information salience, internalization, and persistent effects. In healthcare, analogous risks arise in drug-to-food interactions (e.g., statins alongside grapefruit, which reduces treatment effectiveness), drug-to-disease interactions (e.g., contraindicated medications for hypertensive patients), or device-to-medication interactions (e.g., blood thinners for patients with implanted stents). In line with this broader applicability, our partner firm is developing extensions of the DDI intervention for drug-to-food or drug-to-disease interactions, aiming to enable doctors to internalize and learn these rules and apply them into future decisions.

Beyond healthcare, similar risks also occur in other sectors. In logistics, incompatible goods (e.g., pesticides/hazardous goods paired with grocery items) can compromise safety and quality; in food delivery, risky pairing (e.g., allergens being combined with other deliveries by the same delivery agents) could endanger consumers; and in product development, ingredient incompatibilities (e.g., preservatives that degrade vitamins are added in a multivitamin supplement) reduce efficacy. Across these contexts, our findings highlight that information interventions can help decision-makers not only correct such incompatibilities but also internalize the principles of avoiding harmful pairings, promoting sustained behavioral change and error reduction across diverse operational settings.

A key boundary condition is the completeness and perceived credibility of the information. Alerts perceived as incomplete or unreliable are often ignored, limiting effectiveness (Ancker et al. 2017). In our study, the intervention relies on a comprehensive and trusted DDI database, likely facilitating both corrective actions and proactive learning. This emphasizes the importance of designing interventions with accurate, actionable, and trustworthy information to achieve meaningful behavior change.

Our study has limitations that suggest avenues for future research. First, our findings may not generalize to multi-doctor settings, where additional channels such as peer interactions or informal discussion could alter the effect. Future research could examine how information interventions operate within team-based decision-making contexts for prescription behavior. Second, while our study demonstrates short-term improvements, it is important to assess the long-term impact on both clinical practice and patient outcomes. Extended experimental data could reveal whether these benefits translate into meaningful improvements in patient safety metrics, such as reduced morbidity or mortality. Finally, persistent overrides suggest that complementary strategies may be needed to further reduce prescription errors, such as AI-based decision support and training.

In conclusion, this study suggests practical applications for healthcare providers and policymakers, emphasizing the effectiveness of information interventions in enhancing prescription safety. By fostering learning, such interventions represent a promising tool in reducing healthcare costs and improving patient safety outcomes.

## Endnotes

---

[1] During the first week of the experiment, ~89% of the treatment doctors were exposed to the intervention as they began prescribing interacting drugs. The remaining ~11% of doctors were exposed to the intervention in the subsequent weeks when they started writing interactions. This early and widespread exposure ensured that the experiment was effectively launched within the first week, eliminating the need to consider the treatment as staggered. As a result, there was no necessity to apply methodologies such as staggered difference-in-differences to identify the effects (Baker et al. 2022).

[2] The unequal assignment reflected concerns raised by our partner that exposing too many doctors to the unknown effect of the intervention at the time, such as cognitive overload, potentially result in higher disengagement or increased dropouts. Such attrition could compromise both the integrity of the study and the overall benefits of the platform. Ultimately, however, attrition among doctors was minimal and unrelated to the intervention, as only three doctors from the treatment group who dropped out had stopped prescribing before the experiment began, with their last prescription written 14, 11, and 2 days prior to the launch.

[3] We implement a bootstrapping procedure by resampling individual doctors with replacement, estimating equations (2) and (3) using fixed-effects OLS for each resample. This approach also accounts for the nested nature of the data and computes the bootstrapped standard errors that adjust for correlation between the error terms within cluster (Altman et al. 2021). Because the two estimators of the indirect effect $\alpha_1 \times \beta_3$ are correlated due to sampling error, confidence intervals are calculated based on the empirical distribution of the estimates from each resample (Altman et al. 2021).

[4] The types of molecules prescribed remain largely consistent before versus after the intervention, both in all prescriptions and in uncorrected prescriptions. At the same time, the distribution of common errors (e.g., the top 10 common errors, which account for ~40% of all errors) is stable. Analysis of these common errors shows that our intervention also effectively reduces these frequent errors. These consistent results are omitted for brevity.

[5] The patient volume is the same as prescription volume.

## Online Appendix

### A1. Distribution Plots

### Figure A1. Density and Cumulative Distribution of Variables

(a) *Number of DDIs*

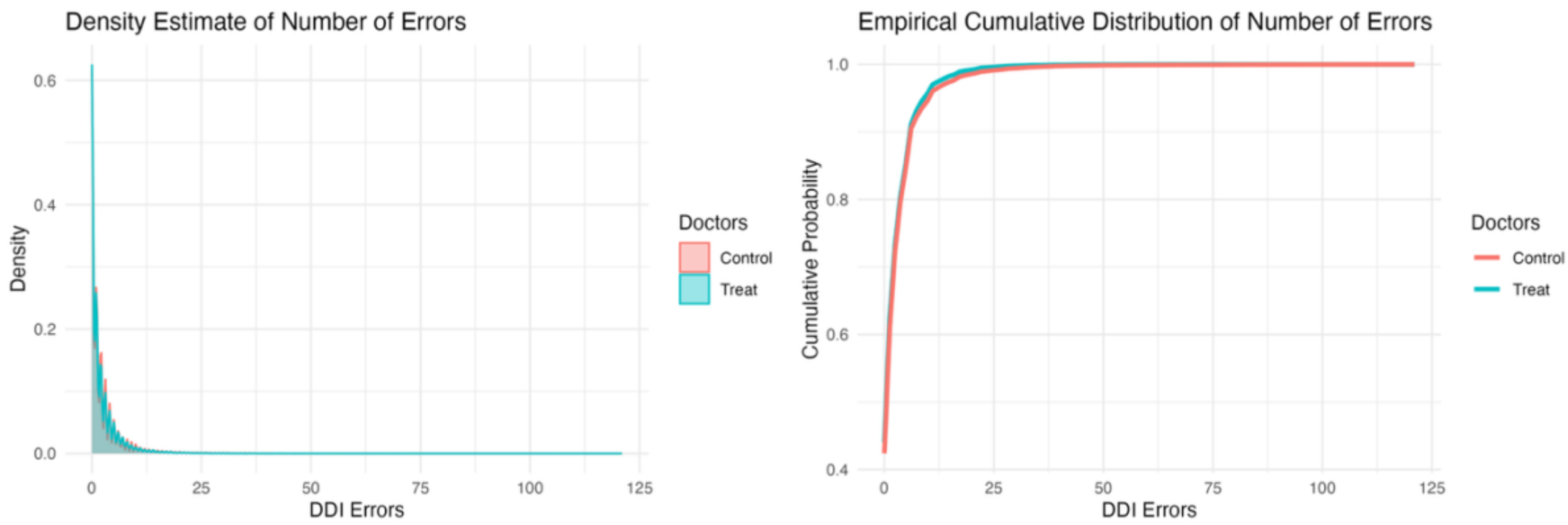

(b) *Number of medicines prescribed*

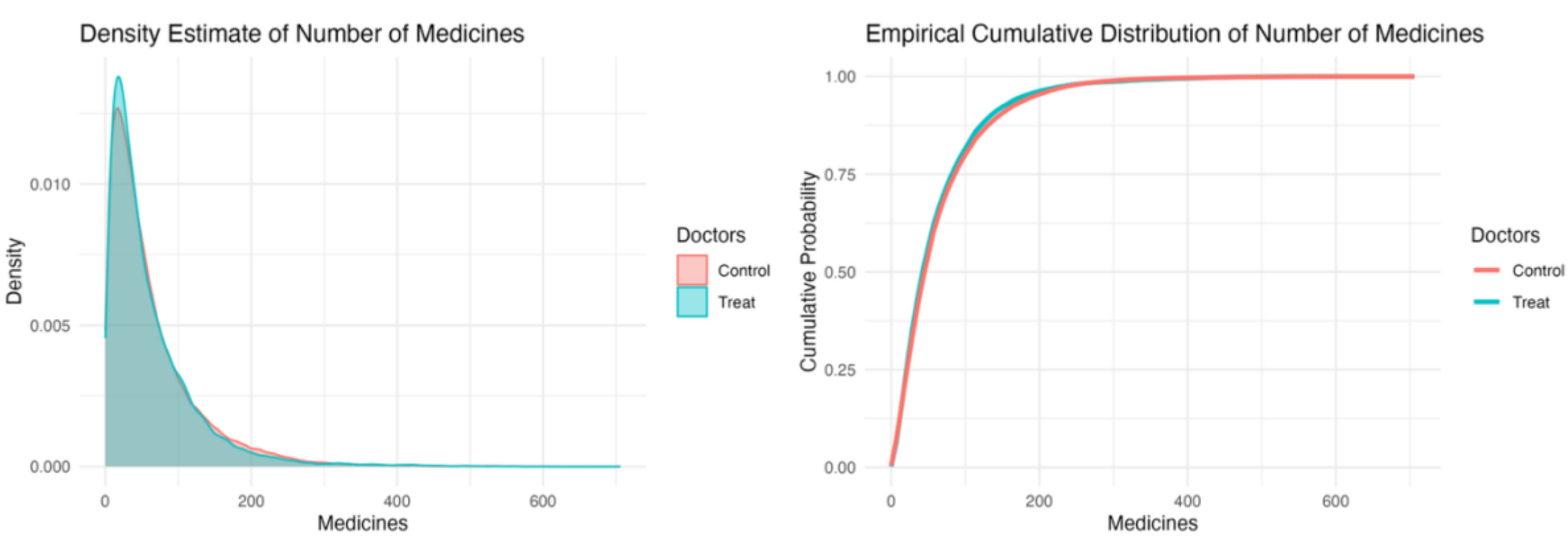

(c) *Total number of active days*

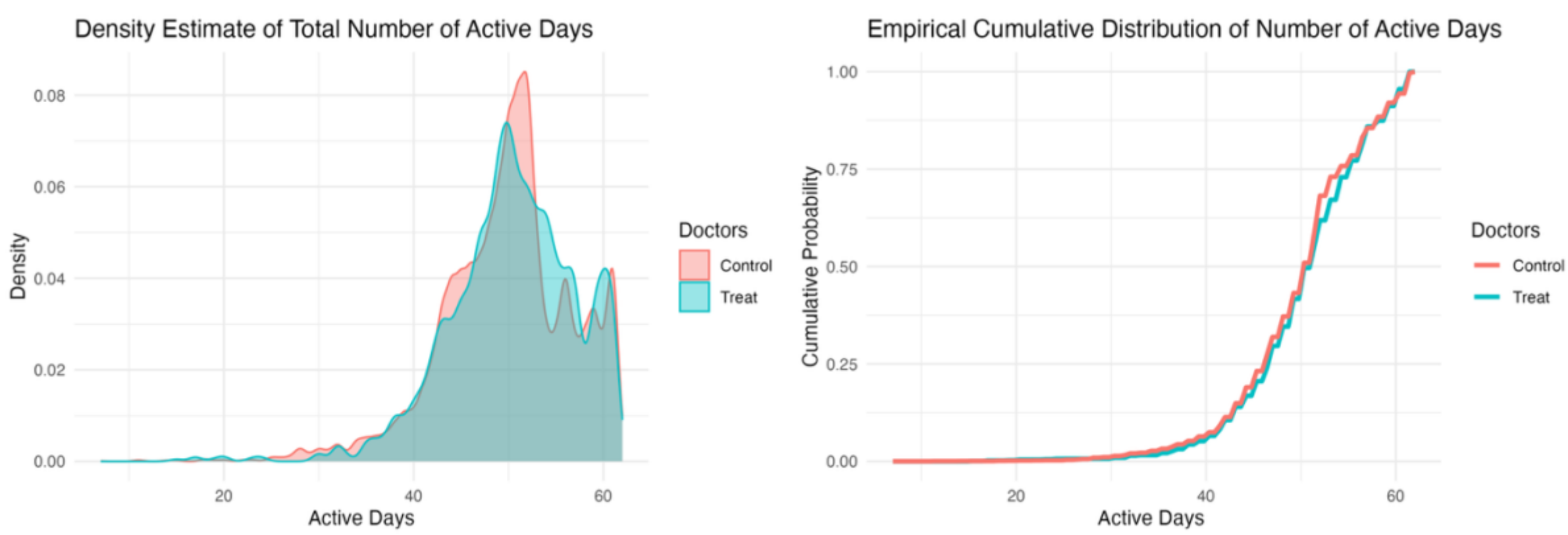

(d) *Total number of patients*

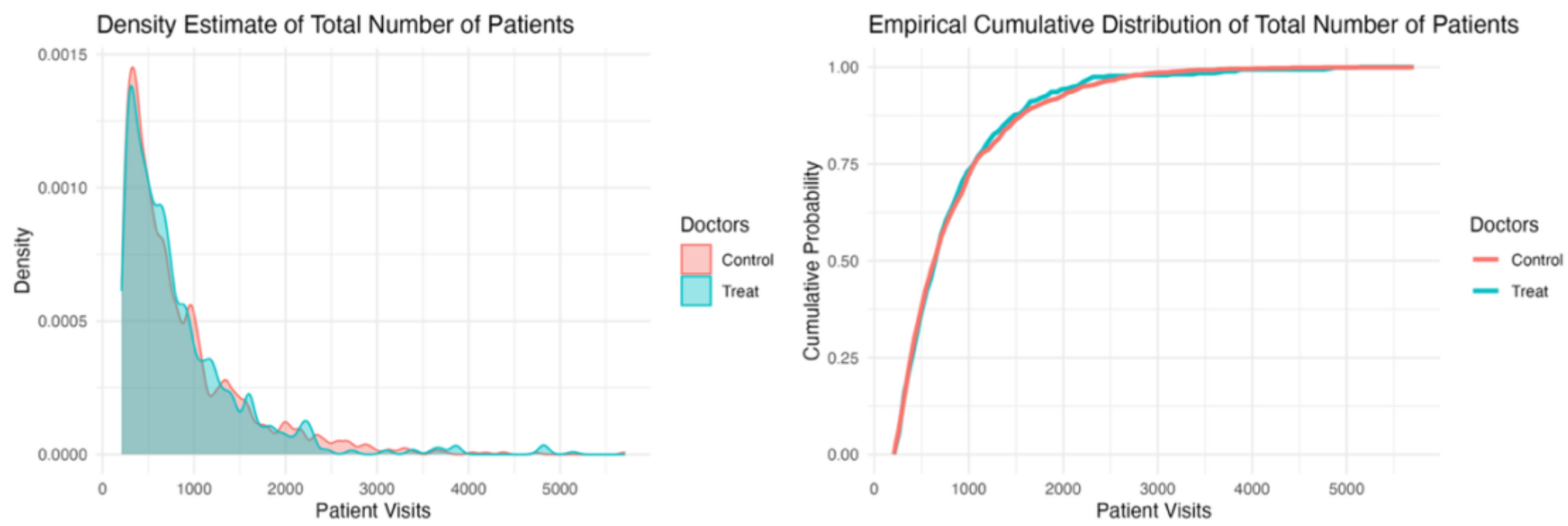

(e) *Number of daily patients*

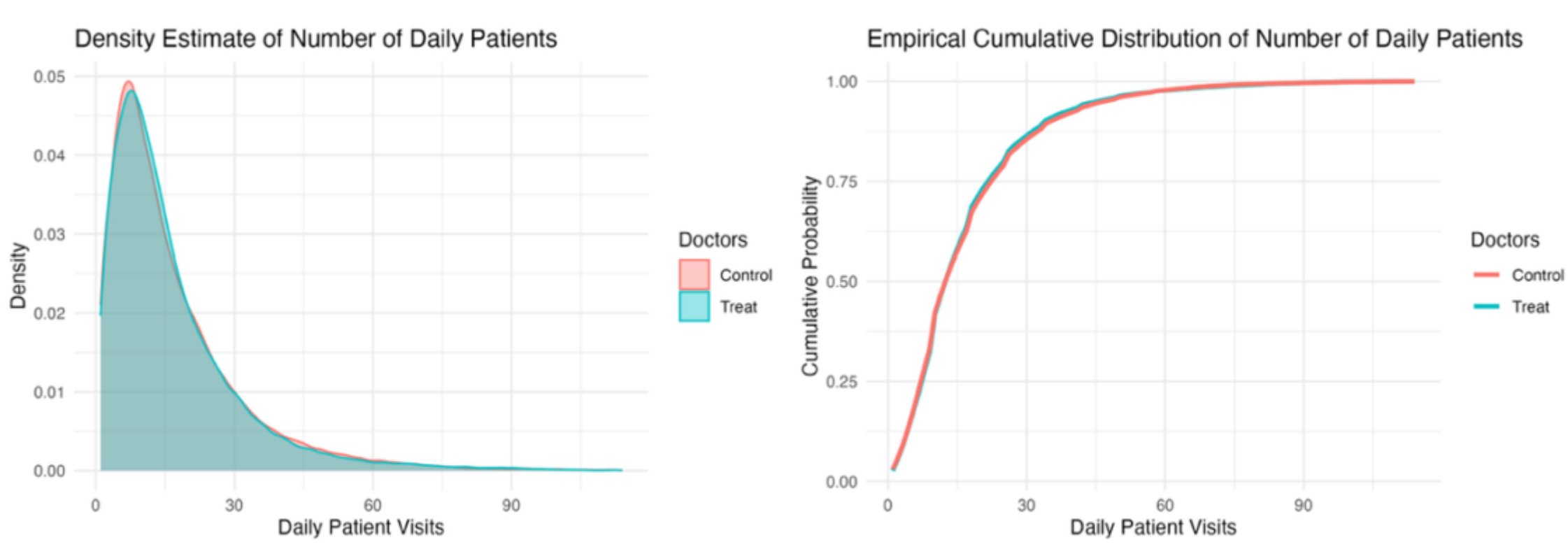

(f) *Number of patient revisits*

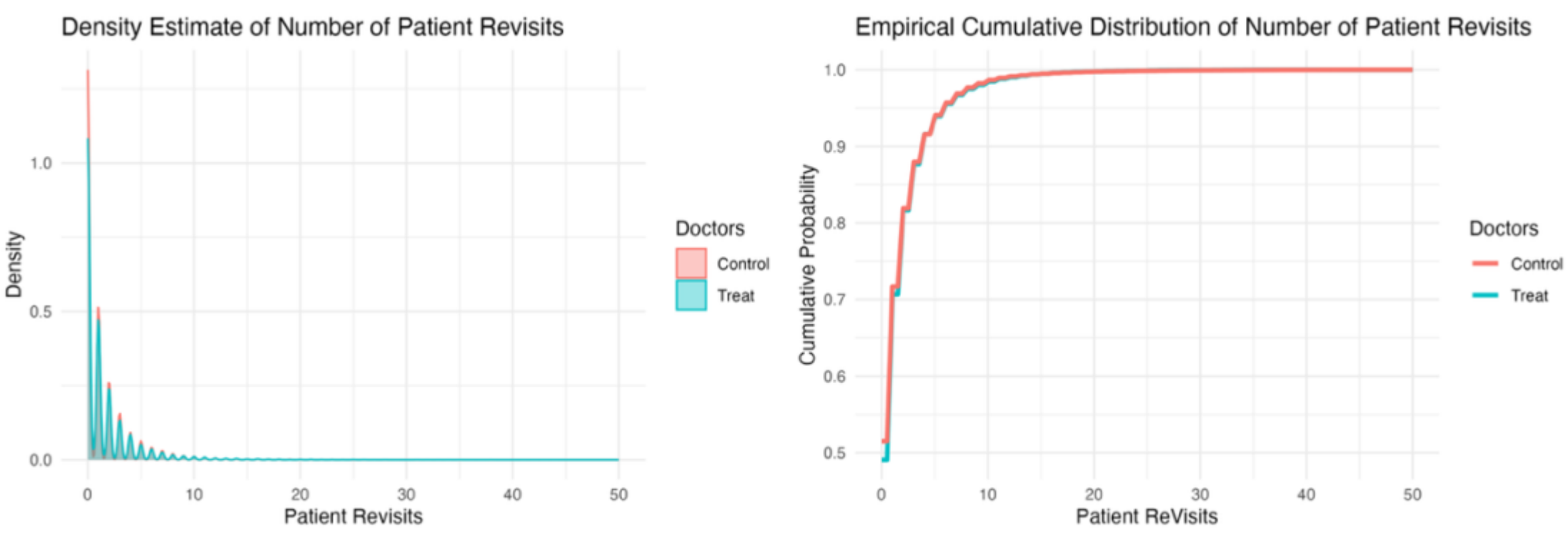

(g) *Experience*

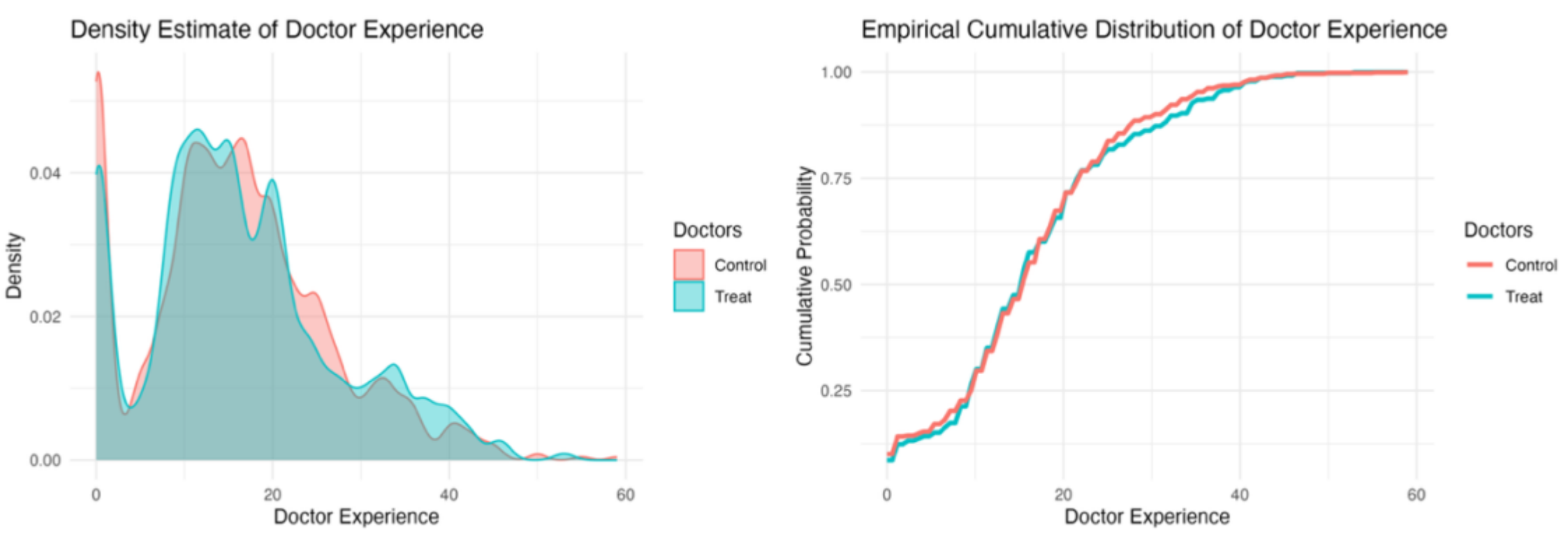

(h) *Specialization* (Binary)

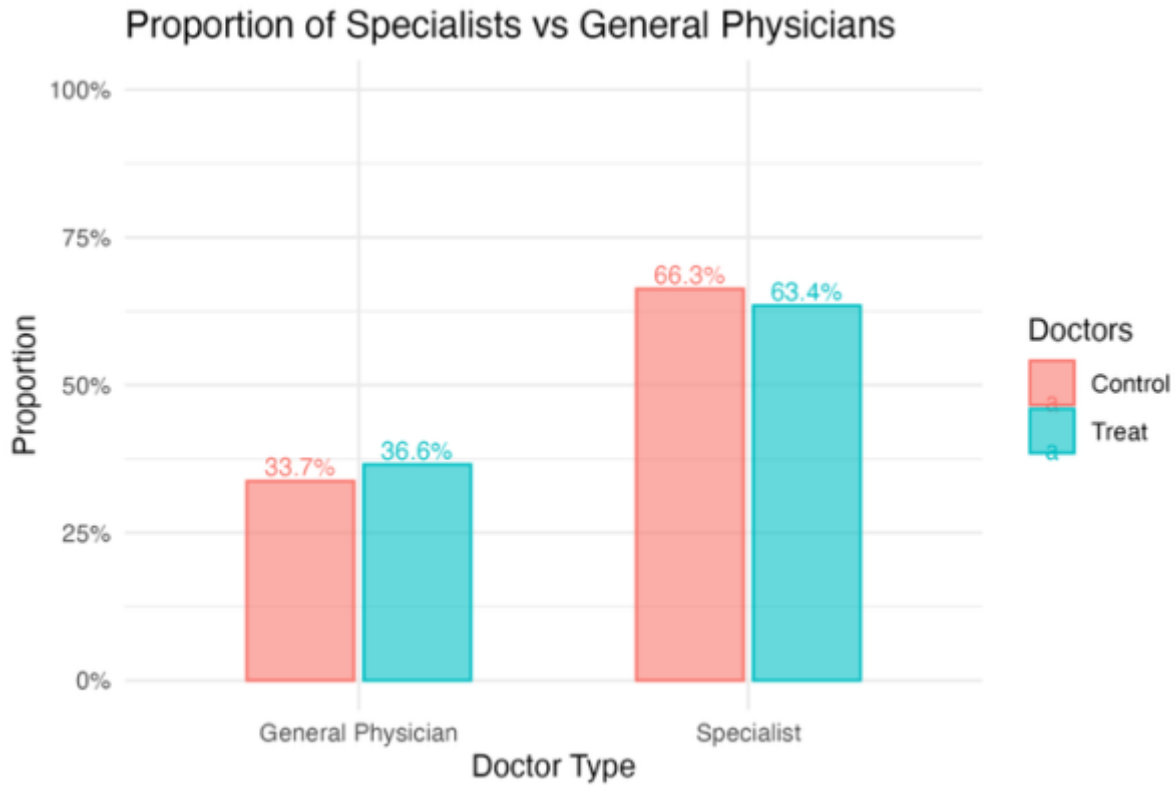

**A2. Parallel Trends**

We use the difference-in-differences (DID) method to estimate the effect of the intervention. A critical identifying assumption of this approach is the existence of parallel trends of the dependent variable between the treatment and control groups during the pre-intervention period. It is assumed that these parallel trends would persist without the intervention during the post-intervention period. Therefore, any changes post-intervention can be attributed to the treatment. While the parallel trend in the post-treatment period is inherently unverifiable, we investigate pre-intervention trends to support the assumption as follows.

We provide statistical evidence to support the parallel trend assumption. Following the methodology of Cui et al. (2022), we use *Week* dummy variables to test whether the treatment and control groups follow the parallel trend during the pre-treatment period. Specifically, we estimate equation A2 to check whether any of the *Week* dummies are statistically significant, which would indicate a violation of the parallel trend assumption.

$$Errors_{it} = \sum_{s=-13}^{-2} \beta_s \times Treatment_i \times Week_s^t + Medicine_{it} + Doctor_i + Week_s^t + \epsilon_{it} \quad \text{(A2)}$$

Here, the week dummy variable $Week_s^t$ where $s \in \{-13, -12, \ldots -2, -1\}$, represents the relative weeks before the start of the experiment, with the final week before the start of the experiment (June 30, 2022) as the reference week ($week_{-1}$). Table A2 shows no significant difference in errors between treatment and control doctors before the experiment, supporting the validity of the parallel trend.

We then provide the visual evidence of the statistic results in Figure A2 accordingly. The weekly estimates of the number of errors between treatment and control groups during the pre-intervention period (relative weeks –13 to –2 of the launch) do not deviate significantly from zero. This pattern suggests that, prior to the intervention, both groups followed a comparable pattern in prescribing errors, thereby validating the parallel trends assumption.

**Table A2. Parallel Trend Test**

| | *Number of errors* |
|---|---|
| $Treatment \times Week_{-13}$ | 0.050 (0.102) |
| $Treatment \times Week_{-12}$ | 0.064 (0.081) |
| $Treatment \times Week_{-11}$ | 0.061 (0.077) |
| $Treatment \times Week_{-10}$ | 0.061 (0.074) |
| $Treatment \times Week_{-9}$ | 0.053 (0.081) |
| $Treatment \times Week_{-8}$ | −0.049 (0.075) |
| $Treatment \times Week_{-7}$ | 0.079 (0.072) |
| $Treatment \times Week_{-6}$ | 0.024 (0.081) |
| $Treatment \times Week_{-5}$ | 0.005 (0.075) |
| $Treatment \times Week_{-4}$ | 0.080 (0.081) |
| $Treatment \times Week_{-3}$ | 0.090 (0.070) |
| $Treatment \times Week_{-2}$ | 0.082 (0.066) |
| $R^2$ Adj. | 0.786 |

*Note: N = 119,070. SEs are clustered at the doctor level. Estimated with all control variables, doctor fixed effect, and week fixed effect. * p < 0.1, ** p < 0.05, *** p < 0.01.*

**Figure A2. Parallel Trend of Errors on a Weekly Basis**

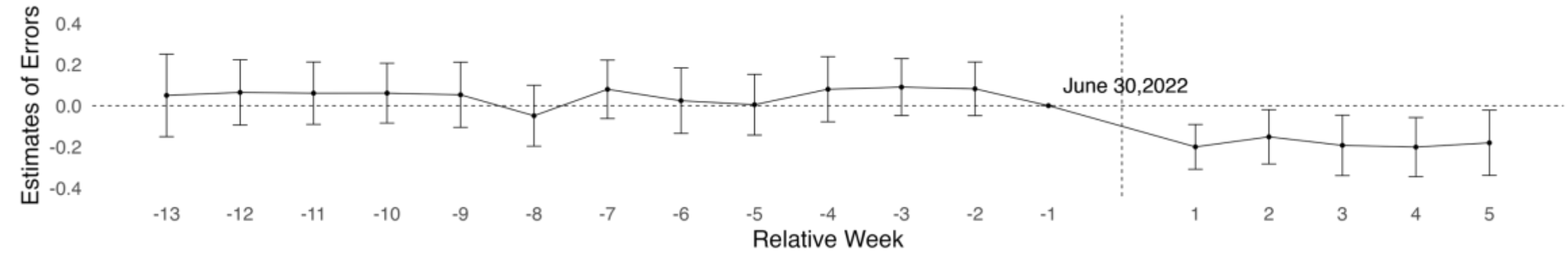

**A3. Estimation of Economic Value and Lives Saved Through the DDI Intervention**

We estimate the cost burden associated with DDI errors and the potential cost-saving impact of our intervention. Due to the limitations in obtaining precise data on the average cost of DDIs, we use the average cost of adverse drug events as a proxy, given that DDIs constitute a major contributor to such events. 1% of hospital admissions resulting from adverse drug reactions are attributed to DDIs (Mousavi and Ghanbari 2017).

Based on this approach, the average cost of hospitalization with an adverse drug event per patient was Rs.4,945 (US$115) in 2008 (Rajakannan et al. 2012). We use equation A3.1 to perform the calculation (noting that the limitation of this estimate is the use of credible data from different years. Our methodology provides a conservative estimate for future years, recognizing that these numbers are bound to increase). Therefore, for India, with reference to the annual hospitalization rate of 3.7% in 2014, and an observed DDI reduction of ~ 8.6% in a one-month experiment period, the estimated annual economic savings from the intervention are approximately US$4.8 million (1,307 million× 3.7% ×1% × 8.6%× $115).

$$Total\ savings\ =\ total\ population \times\ population\ hospitalization\ rate\ \times$$
$$DDIs\ hospitalization\ rate\ \times\ reduction\ in\ DDIs \times cost\ \ per\ incidence \quad (A3.1)$$

Similarly, we estimate the number of lives saved using equation A3.2. Based on a 0.32% annual mortality rate for serious adverse drug reactions among hospitalized patients (Mude and Mohite 2023), the estimated number of lives saved annually in India due to reduced DDIs is approximately 134 (1,307 million× 3.7% ×1% × 8.6%× 0.32%).

$$Total\ live\ save = total\ population \times\ population\ hospitalization\ rate\ \times$$
$$DDIs\ hospitalization\ rate\ \times\ reduction\ in\ DDIs \times death\ rate \quad (A3.2)$$

## A4. Alternative Methods

In addition to the DID approach (ordinary least square), we estimate the effect size of the intervention using the alternative methods outlined below.

### A4.1. Negative Binomial Regression

To account for the count nature of the dependent variable *number of errors* in equation (1), we use negative binomial regression to address potential over-dispersion in the data. Table A3 shows that the number of errors for treatment doctors decreases by 8.9% ($e^{-0.093} - 1$), compared to control doctors. This reduction remains consistent with the trend observed in the main results, demonstrating the robustness of the analysis.

**Table A3. Effect of the Intervention on Errors**

| | *Number of errors* |
|---|---|
| *Treatment× After* | -0.093*** |
| | (0.021) |
| $R^2$ Adj. | 0.261 |

*Note: N = 162,395. SEs are clustered at the doctor level. Estimated with all control variables, doctor fixed effect, and date fixed effect. * p < 0.1, ** p < 0.05, *** p < 0.01.*

### A4.2. Matched Treatment and Control Doctors

To assess the robustness of our findings, we use the nearest neighbor matching without replacement (Abadie and Imbens 2011), matching each treatment doctor with the most similar control doctor based on pre-treatment variables. In column 1, matching implements the same variables used in the balance checks, *number of pre-treatment active days, number of daily pre-treatment patients, number of daily pre-treatment medicines prescribed, number of daily pre-treatment revisits, and doctor specialization*. The *experience* variable is excluded for this matching due to data availability constraints (available for approximately 80% of the observations). Given the larger size of the control group, each treatment unit is matched 1:1 with the closest observation in the control group. This matching process creates a set of doctors who are statistically comparable with an equal sample size prior to the experiment.

In column 2, we incorporate *experience* into the matching process, which reduces the matched sample to approximately 80% of the original size due to missing values. We then conduct DID estimates on the matched samples. As shown in Table A4, the impact of the intervention post-implementation remains statistically significant and robust: -0.209 ($p$<0.01) when excluding experience from the matching and -0.200 ($p$<0.01) when including it. These estimates align with those from the unmatched analysis, reinforcing the internal validity of our results.

**Table A4. Effect of the Intervention on Errors (Matching)**

| | *Number of errors* | |
|---|---|---|
| | *Experience excluded* | *Experience included* |
| *Treatment × after* | -0.209*** | -0.200*** |
| | (0.069) | (0.076) |
| N | 90,616 | 72,246 |
| $R^2$ Adj. | 0.748 | 0.747 |

*Note: Estimated with all control variables, doctor fixed effect, and date fixed effect. * p < 0.1, ** p < 0.05, *** p < 0.01.*

### A4.3. Generalized Synthetic Control (GSC)

We apply a synthetic control method to estimate the intervention effect (Choudhary et al. 2022, Xu 2017). Unlike the DID approach, this method does not rely on the parallel trend assumption. One of the key advantages of the synthetic control method is that it statistically matches the dependent variables, ensuring no significant differences between the treatment and control groups in the pre-treatment period. Additionally, the GSC method reduces the differences in other covariates, enhancing the robustness of the results obtained through the DID. We provide the results in Table A5, which are consistent with our DID estimates.

**Table A5. Effect of the Intervention on Errors (Generalized Synthetic Control)**

| | *Number of errors* |
|---|---|
| *Treatment × After* | -0.155*** |
| | (0.047) |
| N | 202,062 |
| MPSE | 4.761 |

*Note: Estimated with all control variables, doctor fixed effect, and date fixed effect. MPSE reports the mean squared prediction error of the cross-validated model. * p < 0.1, ** p < 0.05, *** p < 0.01.*

## A5. Robustness Checks

In this section, we discuss several tests conducted to assess the robustness of the results.

**A5.1. Estimation with Outliers**

We replicate the analysis of estimating equation (1) on the complete dataset of 162,697 doctor-day observations, including all outliers. Column (2) of Table A6 reports that the number of errors decreases by ~9.3% (= -0.215/2.32) in the treatment group due to the intervention using the full dataset, consistent with the main results obtained in column (1) (reproduced from Table 4 of the manuscript).

**Table A6. Effect of the Intervention on Errors**

| | (1) | (2) |
|---|---|---|
| | *Number of errors* (Outliers excluded) | *Number of errors* (Outliers included) |
| *Treatment× After* | −0.186*** | -0.215*** |
| | (0.052) | (0.059) |
| *Number of medicines* | 0.043*** | 0.047*** |
| | (0.002) | (0.003) |
| N | 162,395 | 162,697 |
| $R^2$ Adj. | 0.786 | 0.789 |

*Note: SEs in parentheses are clustered at the doctor level. Estimated with doctor fixed effect and date fixed effect. * p < 0.1, ** p < 0.05, *** p < 0.01.*

**A5.2. Estimation without Covariates**

We provide our results without controlling for *the number of medicines* in column (2) of Table A7, which remain consistent with our main findings in column (1) (reproduced from Table 4 of the manuscript).

**Table A7. Effect of the Intervention on Errors (With and Without Covariates)**

| | (1) | (2) |
|---|---|---|
| | *Number of errors* (With covariates) | *Number of errors* (Without covariates) |
| *Treatment× After* | −0.186*** | -0.142*** |
| | (0.052) | (0.053) |
| *Number of medicines* | 0.043*** | |
| | (0.002) | |
| $R^2$ Adj. | 0.786 | 0.697 |

*Note: N = 162,395. SEs are clustered at the doctor level. Estimated with doctor fixed effect and date fixed effect. * p < 0.1, ** p < 0.05, *** p < 0.01.*

**A5.3. Error Classification Using 7-Day Moving Window**

Table A8 shows that our results remain robust when using a 7-day moving window to classify repeated and new errors, which captures short-term learning dynamics more effectively. Specifically, repeated errors are defined as those committed by the same doctor within the preceding 7 days, while new errors refer to those not observed during that period. Using this alternative definition, we continue to find significant reductions in both repeated errors (–0.118, $p < 0.01$) and new errors (–0.106, $p < 0.01$), reinforcing the conclusion that the intervention promotes both avoidance of past errors and transfer of learning to novel prescribing situations.

**Table A8. Effect of the Intervention Without Removal (7-day Moving Window)**

| | (1) | (2) |
|---|---|---|
| | *Number of Repeated Errors* | *Number of New Errors* |
| *Treatment × After* | -0.118*** | -0.106*** |
| | (0.044) | (0.016) |
| $R^2$ Adj. | 0.793 | 0.352 |

*Note: N = 152,819. Prescriptions from the first week (4-10 April 2022) are excluded. SEs are clustered at the doctor level. Estimated with all control variables, doctor fixed effect, and date fixed effect. * p < 0.1, ** p < 0.05, *** p < 0.01.*

### A6. Effect Persistency After Eleven Weeks

In Figure A3, the result indicates a consistent reduction in errors within the treatment group from June 30, 2022, to September 14, 2022, suggesting a lasting impact of the intervention.

**Figure A3. Error Prevalence for the Treatment Group**

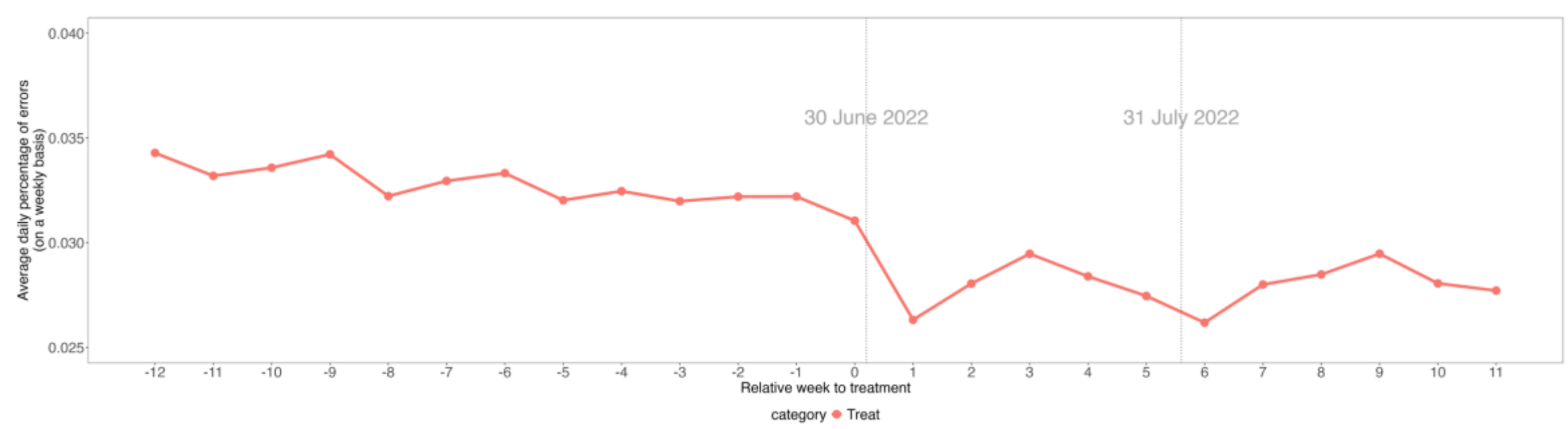